\documentclass[10pt]{article}
\usepackage{amssymb}
\begin{document}

\newtheorem{theorem}{Theorem}
\newtheorem{proposition}{Proposition}
\newtheorem{remark}[theorem]{Remark}
\newenvironment{proof}[1][Proof]{\textbf{#1.} }{\ \rule{0.5em}{0.5em}}


%
\title{\bf Fourier analysis and holomorphic decomposition \\
on the one-sheeted hyperboloid}
\author{Jacques Bros$^1$ and Ugo Moschella$^2$ \\
$^1$Service de Physique Th\'eorique, C.E. Saclay,\\
91191 Gif-sur-Yvette, France \\
$^2$Dipartimento di Scienze Chimiche, Fisiche e Matematiche, \\
Universit\`a dell'Insubria,  Via Valleggio 11, 22100 Como,
Italia\\
and INFN, Sezione di Milano.}

\maketitle

\begin{abstract}
We first prove a Cauchy-type integral representation
for classes of functions holomorphic in
four priviledged tuboid domains of the
quadric $X^{(c)}$ in ${\mathbb C}^{3}$, defined as
the complexification of the one-sheeted hyperboloid  $X$ with equation
$ x_0^2 - x_1^2 - x_2^2 = -1 $. From a physical viewpoint,
this hyperboloid can be used for describing
both the two-dimensional de Sitter and anti-de Sitter universes.

For two of these tuboids,
called ``the Lorentz tuboids'' ${\mathcal T}^+$ and ${\mathcal T}^-$
and relevant for de Sitter Quantum Field Theory,
the boundary values onto $X$ of functions holomorphic in these domains
admit {\it continuous} Fourier-Helgason-type transforms $\tilde f_{\pm, \nu}(\xi)$, where $\nu$ labels
the representations of the principal series of the group $ SO_0(1,2) $ and $\xi$ belongs
to the asymptotic future cone $C^+$ of $X$.
Considering the case of functions invariant under a stabilizer subgroup $SO_0(1,1)$, the link of
the previous transformation with the spherical Laplace transformation of invariant Volterra kernels
is exhibited.
For the other two tuboids, called the ``chiral tuboids''
$T_{\rightarrow}$ and
$T_{\leftarrow}$ and relevant for anti-de Sitter Quantum Field Theory,
the boundary values onto $X$ of functions holomorphic in these domains
admit {\it discrete} Fourier-Helgason-type transforms $\tilde f_{\leftrightarrows,\ell} (\xi) $,
where $\ell$ labels
the representations of the discrete series of the group $ SO_0(1,2) $ and $\xi$ varies in
corresponding domains $C_{\leftrightarrows}$ of the complexified of $C^+$ in ${\mathbb C}^3$.
In both cases, the inversion formulae for these transformations
are derived by using the previous Cauchy representation for the respective classes of functions.
The decomposition of functions on $X$ into sums of boundary values of holomorphic functions
from the previous four tuboids gives
a complete and explicit treatment of the Gelfand-Gindikin
program
for the one-sheeted hyperboloid.

\end{abstract}

\newpage

\section{Introduction}

Although pertaining to the general approach to Fourier analysis on
symmetric spaces $G/K$ whose framework and methods have been
developped in particular by S. Helgason \cite{[H1]} \cite{[H2]},
the case of harmonic analysis on the one-sheeted hyperboloid has
necessitated a special treatment, due to the fact that it exhibits
a situation (in fact, the simplest one) in which $K$ is a {\sl
non-compact} subgroup of $G$. This has been performed in
particular in works by V. Molchanov \cite{[Mol]} and J. Faraut
\cite{[F]}. In this article, we wish to give a new presentation of
the Fourier-Helgason transformation on the one-sheeted hyperboloid
$X$ which is more related to complex analysis on the corresponding
complexified quadric $X^{(c)}$ with equation $z^2 \equiv z_0^2-
z_1^2-z_2^2 \ =\ -1$ in ${\mathbb C}^3$ (considered as a
homogeneous space of the complex Lorentz group $
SO_0(1,2)^{(c)}$); in fact, the results of this study will closely
parallel those of the usual Fourier analysis on ${\mathbb R}^2$ in
connection with complex analysis on ${\mathbb C}^2$.

Our starting point is the existence of four distinguished holomorphy domains
in $X^{(c)}$ which we call ``tuboids'' for the following reasons:

i) there exists a global holomorphic representation of $X^{(c)}$
in ${\mathbb C}^2$ which displays a correspondence between these
four domains and the tubes defined as the products of upper and
lower half-planes in these two complex variables.

ii) each of these domains is bordered by the whole set of real points of
$X^{(c)}$ (i.e. $X$), and is locally a tuboid
in the sense of \cite{[BI]} (see also the Appendix of
\cite{[BM]}): the notion of boundary values of holomorphic
functions from these domains on the reals
is thus well-defined (in the sense of functions or of distributions).

These four domains are invariant under the action of the {\sl
real} group $ SO_0(1,2)$ on $X^{(c)}$. The first two domains
${\mathcal T}^+$ and ${\mathcal T}^-$ play the same role as the
Lorentz tubes $T^{\pm}  = {\mathbb R}^3 + i V^{\pm}$ in ${\mathbb
C}^3$ ($V^+$ and $V^-$ denoting the open future and past cones in
${\mathbb R}^3$); they will be called below ``Lorentz tuboids of
$X^{(c)}$''. The other two domains ${\mathcal T}_\rightarrow$ and
${\mathcal T}_\leftarrow$ which are not simply-connected and are
distinguished from each other by an orientation prescription will
be called ``chiral tuboids of $X^{(c)}$''.

We shall consider classes of holomorphic functions
in these four tuboids, sufficiently regular at infinity so as to admit a
Cauchy-type integral representation in terms of their boundary values on $X$.
The corresponding Cauchy kernel will be seen to be
proportional to the inverse of
the Minkowskian quadratic form $(z-z')^2$ restricted to $X^{(c)} \times X^{(c)}$;
note that this kernel is invariant under the action of the complexified group
$ SO_0(1,2)^{(c)}$ of $SO_0(1,2)$ on $X^{(c)}$.

\vskip 0.2cm Our purpose is to provide {\sl a theory of the
Fourier-Helgason (FH) transformation specifically adapted to the
classes of functions on $X$ which are boundary values of
holomorphic functions from either one of these four tuboids.}
While this paper is purely mathematical, its results have a
natural physical interpretation in the context of de Sitter and
anti-de Sitter Quantum Field Theory, where the relevant
correlation functions belong precisely to such classes
\cite{[BM],[BEMDS],[BBMS],[BEMADS]}.

\vskip 0.2cm Introducing the bilinear form $[z\cdot z'] = z_0 z'_0
-z_1 z'_1 - z_2 z'_2 $ on ${\mathbb C}^3$, we shall make use of
the FH-kernel $[z\cdot \xi]^s$ for various types of configurations
of the pair $(z,\xi)$, with $z^2 \equiv [z\cdot z] = -1$  and
$\xi^2 \equiv [\xi\cdot \xi] =0$.

In the case of holomorphic functions in the Lorentz tuboids
(already considered in \cite{[BM]}), one will take advantage of
the fact that the complex number $[z\cdot \xi]$ remains in the
upper (resp. lower) half-plane when $z$ varies in ${\mathcal T}^+$
(resp. ${\mathcal T}^-$) and  $\xi$ lies in the asymptotic future
cone $C^+ = \partial V^+$ of $X$. One then makes use of the
following form of the FH-kernel $[x_{\pm}\cdot\xi]^s = \lim_{z \in
{\mathcal T}^{\pm}, {\rm Im}z \to 0} [z\cdot\xi]^s $ in place of
the Gelfand-type forms ${| [x\cdot\xi]|}^s $ and ${\rm
sgn}([x\cdot\xi]){| [x\cdot\xi]|}^s $ currently used for
introducing the FH-transformation on $X$ (see e.g. \cite{[F]}
\cite{[Mol]} and references therein). In fact, it will be seen
that for the functions $f(x)$ which are boundary values of a
holomorphic function respectively from ${\mathcal T}^+$ or
${\mathcal T}^-$ our prescription for $[x\cdot \xi]^s$ selects a
\it unique non-vanishing \rm FH-transform, denoted respectively
$\tilde f_{-,\nu} (\xi)$ or $\tilde f_{+,\nu} (\xi)$, having
restricted $s$ to vary in the range of values $s = -1/2- i \nu$
which label the principal series of irreducible unitary
representations of the group $ SO_0(1,2)$. The existence of such a
unique relevant FH-type transform is the analogue of \it the
support property of the Fourier transforms \rm for the functions
(or tempered distributions) in ${\mathbb R}^3$ which are boundary
values of holomorphic functions in either tubes $T^+$ or $T^-$ of
${\mathbb C}^3$: such functions are characterized by the fact that
their Fourier transforms have their support contained in the
closure of either one of the cones $V^{\pm}$ (see \cite{[Sch]},
chapter 8).

In order to invert this FH-transformation,
we shall also make use of the analyticity properties in $z$ of the inverse FH-kernel
$[z\cdot\xi]^{-1/2 + i \nu}$ which allow one to define the expected inverse as a holomorphic function
$F(z)$ in the relevant Lorentz tuboid ${\mathcal T}^{\pm}$ of $X^{(c)}$. The proof that the initially
given function $f(x)$ is indeed the boundary value of $F(z)$ from its tuboid is obtained
by directly computing the composition of the direct and inverse FH-kernels
and showing that it yields explicitly the Cauchy-type representation in
${\mathcal T}^{\pm}$
established at first.
This generalizes the procedure according to which
the standard Cauchy kernel $ (z - x)^{-1}$ emerges from the
Fourier inversion computation \it with support properties \rm, as being equal
(for $z $ in the upper half-plane) to the integral
$ 1/i \int_0^{\infty} e^{i(z-x)t} \ dt$.

Among the functions or distributions $f(x) $ on $X$ which are
boundary values of holomorphic functions from ${\mathcal T}^+$ or
${\mathcal T}^-$ it is interesting to consider those which are
moreover invariant under a subgroup of $SO_0(1,2)$ such as the
stabilizer (isomorphic to $SO_0(1,1)$) of a given ``base point''
$b$ of $X$ (chosen below as $ b = (0,0,1)$). In fact, such
functions or distributions can be identified with invariant
kernels on $X$ which are boundary values of holomorphic functions
defined in the ``cut-domain'' $\{(z,z') \in {X^{(c)} \times
X^{(c)}}; \ (z-z')^2 \in {\mathbb C} \backslash {\mathbb R}^+ \}$.
These objects have been studied under the name of \it perikernels
\rm in \cite{[BV]}, \cite{[B]} and their discontinuities on the
cut $\{(z,z');\ (z-z')^2 \ge 0\} $ define  \it Volterra kernels
\rm in the sense of \cite{[Fa]}. By applying the previous
Fourier-Helgason transformation to this class of
$SO_0(1,1)$-invariant functions and to their discontinuities, we
shall then reobtain the theory of the {\it ``spherical Laplace
transformation'' of invariant Volterra kernels}  of \cite{[FV]},
\cite{[Fa]}.

In the case of holomorphic functions in the chiral tuboids, the
definition of the FH-transformation will be qualitatively
different, since the variable $\xi$ may now vary in two
corresponding domains of the complexified cone $C^{(c)}$ of $C^+$,
while the exponent $s$ is restricted to the set of integral values
$s = -\ell-1$, where $\ell \geq 0$ labels the discrete series of
irreducible unitary representations of $SO_0(1,2)$. This
discretization, which is due to the topological equivalence of
${\mathcal T}_\rightarrow$ and ${\mathcal T}_\leftarrow$ with the
product of $S_1$ with wedges in ${\mathbb R}^3$, would be lifted
if these chiral tubes were replaced by their universal coverings
(i.e. $S_1$ by ${\mathbb R}$), and $X$ and its symmetry group
$SO_0(1,2)$ by their corresponding coverings \cite{[BBMS]}.

After having introduced the geometry of the four tuboids
${\mathcal T}^+$, ${\mathcal T}^-$,
${\mathcal T}_\rightarrow$ and
${\mathcal T}_\leftarrow$
of the complex quadric $X^{(c)}$  in our section 2, we shall devote section 3  to
the proof of a Cauchy-type representation for functions holomorphic in these four domains.

A ``Lorentzian'' FH-transformation is then introduced and studied
in section 4 .For the functions on $X$ which are
boundary values of holomorphic functions in the tuboids ${\mathcal T}^{\pm}$,
the inversion of this FH-transformation is then performed by the
Cauchy kernel method, as explained above.

The case of $SO_0(1,1)-$invariant functions (i.e. of invariant perikernels)
and its connection with the theory of spherical Laplace transformation
is treated in section 5.

Section 6 is devoted to the ``chiral'' FH-transformation and to
the corresponding representation of the functions on $X$ which are boundary
values of holomorphic functions in the tuboids
${\mathcal T}_\rightarrow$ and
${\mathcal T}_\leftarrow$.

In the final section 7, we shall summarize our results concerning the decomposition of
functions on $X$ as sums of boundary values of holomorphic functions from the four tuboids
${\mathcal T}^+$, ${\mathcal T}^-$,
${\mathcal T}_\rightarrow$ and
${\mathcal T}_\leftarrow$ of $X^{(c)}$.
Our characterization of this decomposition in terms of {\sl
Lorentzian and chiral FH-transforms of the
four components} gives a complete and explicit treatment of the {\sl Gelfand-Gindikin
program} (see  \cite{[G]}, \cite{[O]} and references therein)
for the case of the one-sheeted hyperboloid in dimension 2.

Extensions of our results in two directions will be given in
forthcoming papers: on the one hand, results concerning
holomorphic functions in the Lorentz tuboids can be generalized to
the case of the complexified $n-$dimensional one-sheeted
hyperboloid in ${\mathbb C}^{n+1}$; on the other hand, results
concerning holomorphic functions in chiral tuboids can be
generalized to the case of the complexified $n-$dimensional
quadric $[z\cdot z] = 1$, where $[z\cdot z]$ denotes a quadratic
form with signature $(+,+,-,\cdots,-)$. These two cases have
respective applications to Quantum Field Theory in de Sitter and
anti-de Sitter spacetime manifolds in dimension $n$ (see
\cite{[BM],[BEMDS],[BBMS],[BEMADS]}).

\vskip 1cm

\section{The four basic tuboids of the complexified one-sheeted hyperboloid}
The space ${{\mathbb R}^3}$ of the variables $x = (x_0, x_1, x_2)$
is equipped with the Minkowskian bilinear form
\begin{equation}
[x\cdot y] =  x_0 y_0 - x_1 y_1 - x_2 y_2,
\ \ \ \ (x^2 = [x\cdot x]).
\label{product}
\end{equation}
We introduce the ``light-cone''
\begin{equation}
C = \left\{x\in {{\mathbb R}}^{3}: \ x^2 =0;\ x\neq 0 \right\},
\end{equation}
the (closed) ``future cone''
\begin{equation}
\overline V^{+} = \left\{x\in {{\mathbb R}}^{3}: \ x^2 \geq 0,\
{x_0}\geq 0 \right\}
\end{equation}
its interior $V^{+} $ and its boundary $C^+ = C \cap \overline V^+$. One
defines similarly $V^- = -  V^+$, etc...

The one-sheeted hyperboloid ${X} = \left\{x \in {{\mathbb R}^3}:
{x^2} = - 1\right\}$ is equipped with a {\em ``causal'' ordering
relation} induced by that of the ambient Minkowskian space
${{\mathbb R}}^{3}$, namely $\forall (x,y) \in X\times X,\;\;x\geq
y\; \leftrightarrow \; x-y \in \overline V^{+}$. The ``future
cone'' \footnote{This terminology is justified by the
interpretation of the quadric $X$ as a two-dimensional de Sitter
spacetime.} of a given point $x$ in $X$ is $\Gamma^+{(x)}= \{y \in
X: y \geq x \; \}$. Analogously one defines $\Gamma^-{(x)}$. The
``light-cone''$ {\partial \Gamma}(x)$ of $x$ on $X$, namely the
boundary set of $\Gamma^+{(x)}\cup\Gamma^-{(x)}$,  is the pair of
linear generatrices of $X$ containing the point $x$.

The invariance group of $X$ is the pseudo-orthogonal group
$SO(1,2)$ leaving invariant the bilinear form (\ref{product})
and we shall call ``Lorentz group $G$'' the connected component $SO_0(1,2)$ of the latter.
We choose the integration measure ${\rm d}\sigma(x)$ on $X$ as associated with
the $G$-invariant volume form
$ \left.\frac{{\rm d}x_0 \wedge {\rm d}x_1 \wedge {\rm d}x_2}
{{\rm d}(x^2 + 1)}\right|_{X}  $
(in Leray's
notations \cite{[Le]}), namely:
\begin{equation}
{\rm d}\sigma(x)=\frac{ {\rm d}x_1  {\rm d}x_2} {2 |x_0|}
\label{leray}
\end{equation}
Since the group $G$ acts in a transitive way on $X$, it is
convenient to distinguish a {\em base point} $b$ in $X$ which we choose to be
$b=(0,0,1)$.

The complexified hyperboloid $X^{(c)}= \{z = x+iy \in {{\mathbb
C}}^{3}: z^2 =\ -1 \}$ is equivalently described as the set
\begin{equation}
X^{(c)}= \{(x,y)\in {\mathbb R}^{3} \times {\mathbb R}^{3}:\
{x}^{2}-{y}^{2}= -1,\  [x\cdot y] = 0\}. \label{hypec1}
\end{equation}
The set of complex points of $X^{(c)}$ can be decomposed as the union
of the following three disjoint sets
denoted by ${\mathcal T}_L$,
${\mathcal T}_0$ and
${\mathcal T}_{\chi}$, separately invariant under the action of $G$ (defined on $X^{(c)}$ by
$gz = gx + i gy$ for all $g \in G$):

i)
\begin{equation}
{\mathcal T}_0 = \{z= x+iy \in
X^{(c)};\ \  y^2 = x^2 +1 =  0 \}
\end{equation}
is the set of all complex points of the straight lines which generate $X$;

ii)
\begin{equation}
{\mathcal T}_L = \{z= x+iy \in
X^{(c)};\ \  y^2 = x^2 +1 > 0 \}
\label{TL}
\end{equation}
In view of the equation $[x\cdot y] =0$ (\ref{hypec1}), all the points in
${\mathcal T}_L$ are also such that $-1 < x^2 < 0$ and correspondingly $0 < y^2 < 1$.

iii)
\begin{equation}
{\mathcal T}_{\chi} = \{z= x+iy \in
X^{(c)};\ \  y^2 = x^2 +1 < 0\}
\label{Tell}
\end{equation}

These three sets can also be characterized in terms of the three families of planes $\Pi$ containing the
origin and respectively tangent, transverse, or exterior to the light-cone $C$.
While
${\mathcal T}_0$ is the union of complex points of sections of $X$ by planes $\Pi$ tangent to $C$,
${\mathcal T}_L$ appears as the union of complex points $z$ of all (hyperbolic)
sections of $X$ by planes $\Pi$ transverse to $C$ and
${\mathcal T}_{\chi}$ as the union of complex points of all (elliptic)
sections of $X$ by planes $\Pi$ exterior to $C$. In fact, each
complex point $z=x+iy$ of $X^{(c)}$
determines a unique plane $\Pi = \Pi(z)$ containing $x$ and $y$, whose complexified contains $z$ and
$\overline z$, and which belongs to either one of
the three previous categories according to whether
$y^2$ is equal to zero, positive or negative: this is because the two orthogonal vectors
$x$ and $y$ (in the sense of the Minkowskian scalar product) indicate
the type of metric obtained on the plane $\Pi$, which is respectively, in view
of the signs of $x^2$ and $y^2$, either degenerate or Minkowskian or Euclidean.

\subsection{ The set
${\mathcal T}_L$ and the
tuboids ${\mathcal T}^+$ and ${\mathcal T}^-$}

We define the
Lorentz invariant sets ${\mathcal T}^+$ and ${\mathcal T}^-$
as the connected components of
${\mathcal T}_L$,
specified by adding the respective conditions
$y_0 > 0$ and
$y_0 < 0$  to the definition of
${\mathcal T}_L$.

These two sets, which we call ``Lorentz tuboids'' are complex conjugate of each other and
an equivalent definition of them is:
\begin{equation}
{\mathcal T}^{+} = \makebox{\rm T}^+\cap X^{(c)},\;\;\;\;\;
{\mathcal T}^{-} = \makebox{\rm T}^-\cap X^{(c)},
\label{tubi1}
\end{equation}
 where {\rm T}$^{\pm} =
 {\mathbb R}^{3} + i {V^{\pm}}$ are the Lorentz
 tubes
in the ambient space ${{\mathbb C}}^{3}$.

\vskip 0.4cm
\noindent
We shall now give two alternative characterizations of ${\mathcal T}^{\pm}$:
\begin{proposition}
${\mathcal T}^+$ (resp. ${\mathcal T}^-$) is the set of all points $z =x + iy$ in $X^{(c)}$
such that the inequality ${\rm Im}([z\cdot \xi]) >0$ (resp. $<0$) holds for {\rm all} $\xi \in C^+$.
\end{proposition}
The proof is immediate since the condition $\forall \xi \in C^+,\  [y\cdot \xi]>0$ (resp. $<0$) is
equivalent to the condition $y \in V^+$ (resp. $V^-$).

\begin{proposition}
The domains ${\mathcal T}^+$
and ${\mathcal T}^- $
are generated respectively by the action of the
Lorentz group $G$ on the following
one-dimensional subsets (namely
half-circles)
of $X^{(c)}$:
\begin{equation}
\gamma^+  =\ \{ z = x + iy \in X^{(c)}; \, z=  z_u = (i\sin u,0  ,\cos u),\  \, 0<u<\pi \}
\end{equation}
and
\begin{equation}
\gamma^-  =\ \{ z = x + iy \in X^{(c)}; \, z=  z_u = (i\sin u,0  ,\cos u),\  \, -\pi<u<0 \}
\end{equation}
\end{proposition}
\noindent
\begin{proof}
It is sufficient to consider the case of
${\mathcal T}^+$
and $\gamma^+$.
Let $G_b$ be the stabilizer of $b$ in G.
The action of the one-parameter subgroup $G_b$ on $\gamma^+$ generates the set
\begin{equation}
G_b\  \gamma^+  = \{ z  \in X^{(c)}; \,
z= (i\sin u \cosh \alpha,\ i\sin u \sinh \alpha,\  \cos u),\  \, 0<u<\pi\}
\label{Gbgamma}
\end{equation}
Let now $z= x+iy$ be an arbitrary point in ${\mathcal T}^+$; in view of
(\ref{TL}), $x$ belongs to an hyperboloid of the form $x^2 = - \cos^2 u $,
and therefore  (by transitivity) there exists a $g \in G$ such that
$gx = b_u = { \cos u}\  b$.
Then the point $gy$ must be such that $[gy\cdot b_u] = [y\cdot x] =0$
with $(gy)^2=\sin^2 u$ and $(gy)_0 >0 $,
which implies that $gz = b_u + i gy$ is of the form (\ref{Gbgamma});
therefore one has $gz \in G_b\  \gamma^+$ and $z \in G \gamma^+$.
\end{proof}

The domain ${\mathcal T}^+$ is a tuboid over $X$ in $X^{(c)}$
whose profile at the base point $b$ is the cone $V^+ (b) = \{y\in
{\mathbb R}^3; y_{0}> |y_{1}|,\ y_{2}=0 \}$; the latter is in fact
obtained by taking the limit $u \to 0$ from the set $ G_b\
\gamma^+$ into the tangent plane to $X$ at $b$. The domain
${\mathcal T}^-$ is similarly described by replacing the condition
$0<u<\pi$ by $-\pi < u <0$ in the previous analysis: it is a
tuboid whose profile at the point $b$ is the cone $V^-(b) = -
V^+(b)$.

\vskip 0.3cm
\noindent

The following property holds:
\begin{proposition}
The projection of the domain ${\mathcal T}^+$ (or ${\mathcal
T}^-$) in the complex plane of the coordinate $z_2 = -[z\cdot b]$
is the cut-plane $\Theta_{L} = {\mathbb C} \setminus
\{[-\infty,-1] \cup [1,+\infty]\}. $ The image of the domain
${\mathcal T}^- \times {\mathcal T}^+$ (or ${\mathcal T}^+ \times
{\mathcal T}^-$) by the mapping $(z,z') \rightarrow [z\cdot z']$
is the cut-plane $\hat\Theta_{L} = {\mathbb C} \setminus [-\infty,
-1]$.
\end{proposition}
\begin{proof}
a) We first notice that the set ${\mathcal T}_0^+ = \{z=
(i\sin(u+iv),\  0,\ \cos(u+iv)); 0<u<\pi, v \in {\mathbb R} \}$ is
contained in ${\mathcal T}^+$, since all these points $z=x+iy$ are
such that  $y_0 = \sin u \cosh v >0$ and $y^2 = \sin^2 u >0$. One
then checks that the corresponding range of the projection $\{z_2
= \cos (u+iv);\ 0<u<\pi,\ v \in {\mathbb R} \}$ is already
$\Theta_{L}$.

Let us now show that the points exterior to $\Theta_{L}$, namely the points
$z_2$ which are real and such that $|z_2| \geq 1$ cannot be the projections of points $z$ in
${\mathcal T}^+$ or in ${\mathcal T}^-$. In fact, all points $z =(z_0,z_1,z_2) \in X^{(c)}$ with
$z_2^2 >1 $ belong to a complex hyperbola $h(z_2)$
with equation $ {\hat z}^2 \equiv z_0^2 - z_1^2 = z_2^2 -1  >0$.
Keeping the same notations for the two-dimensional Minkowskian bilinear form, we can say
that all the complex points $z= (\hat z, z_2)$,
with $\hat z =(z_0,z_1) = \hat x + i\hat y$ in $h(z_2)$ satisfy the
conditions ${\hat x}^2 - {\hat y}^2 >0$ and $ [{\hat x}\cdot {\hat y}] =0$; the latter imply
${\hat x}^2 >0$ and ${\hat y}^2 \equiv y^2 <0$ and therefore $ z \notin {\mathcal T}_{L}$.
As for the points $z$ such that $z_2^2 =1$, they are such that ${\hat x}^2 = y^2 =0$ and
therefore also not in ${\mathcal T}_{L}$.

b) We first check that the image of the subset $\left\{ \left(
z',z\right) ;z'\in \gamma^- ,z \in {\mathcal T}_0^+ \right\} $ of
${\mathcal T}^- \times {\mathcal T}^+$ in the plane of the
variable $[z \cdot z']$ is already $\hat\Theta_{L} .$ Let in fact:
$z' = ( -i \sin u',0 , \cos u'),$  with $  0< u' <\pi$, and $z \in
{\mathcal T}_0^+ $ as described in a). We then have: $ [z\cdot z']
= - \cos (u+u' +iv)$ with $u+u',v$ varying in the range $0< u+u' <
2\pi;\  v \in {\mathbb R}$, and therefore the corresponding range
of $[z \cdot z']$ is $\hat\Theta_{L} .$

It remains to show that one cannot have $[z\cdot z']$ real and $
\leq -1$, or equivalently $(z-z')^2 \geq 0$, if $z$ belongs to $
{\mathcal T}^+$ and $z' $ to $ {\mathcal T}^-$.  This follows from
a simple argument in the ambient space ${\mathbb C}^3$: the
conditions $z \in {\mathcal T}^+$ and $z'\in {\mathcal T}^-$
imply that the vector $Z= z-z' =X+iY$ belongs to the tube $T^+$,
so that one has $Y^2 >0 , Y_0 >0$. But the condition $Z^2 \geq 0$,
which is equivalent to the pair $X^2- Y^2 \geq 0,\  [X\cdot Y] =0$
cannot be implemented with $Y^2 >0 $, since the latter implies
$X^2 \leq 0$.

It is clear that at each step of this proof, the roles of the tubes ${\mathcal T}^+$ and
${\mathcal T}^-$ can be inverted without changes in the conclusions.
\end{proof}

\subsection{The set ${\mathcal T}_{\chi}$ and
the tuboids ${\mathcal T}_\rightarrow$ and
${\mathcal T}_\leftarrow$ }

We shall also use later the following alternative definition of ${\mathcal T}_{\chi}$:

\begin{proposition}
${\mathcal T}_{\chi}$ is the set of all complex points $z=x+iy$ in $X^{(c)}$
such that there exist two distinct vectors $\xi_+$ and $\xi_-$ in $C^+$
(depending on $y$) satisfying
the conditions:
${\rm Im}([z\cdot\xi_+]) =
{\rm Im}([z\cdot\xi_-]) =  0.$
\end{proposition}
\begin{proof}
The stated conditions are equivalent to the
existence of two distinct planes tangent to the cone $C^+$ and intersecting each other along the
support of $y$. The latter condition is of course equivalent to the fact that $y$ is outside
the union of $V^+$ and $V^-$, namely that $y^2$ is negative.
\end{proof}

\vskip 0.3cm
Let $e$ be any vector in the cone
$V^+$;
for every $z=x+iy$ in
$ {\mathcal T}_{\chi}$,
we put $\epsilon(z) = \makebox{sgn Det}(e,x,y)$, this sign being independent of the choice
of $e$ since the plane $\Pi(z)$ is exterior to $C$. It is then clear that for every $g \in G$,
one has $\epsilon(gz) = \epsilon(z)$, so that the following two sets are Lorentz-invariant.
\begin{equation}
{\mathcal T}_\rightarrow =\{ z = x + iy \in X^{(c)}; \, y^2<0,\, \epsilon(z) =  -  \},
\end{equation}
\begin{equation}
{\mathcal T}_\leftarrow =\{ z = x + iy \in X^{(c)}; \, y^2<0,\, \epsilon(z) = + \}.
\label{chirality}
\end{equation}
These two connected components of
${\mathcal T}_{\chi}$, which we call ``chiral tuboids'',
are complex conjugate of each other.

\begin{proposition}
The domains ${\mathcal T}_\rightarrow $
and ${\mathcal T}_\leftarrow $
are generated respectively by the action of the
Lorentz group $G$ on the following
one-dimensional subsets (namely
half-branches of hyperbola)
of $X^{(c)}$:
\begin{equation}
h_\rightarrow =\ \{ z = x + iy \in X^{(c)}; \, z=  z_v = (0,i\sinh  v,\cosh  v),\  \, v>0\}
\end{equation}
and
\begin{equation}
h_\leftarrow =\ \{ z = x + iy \in X^{(c)}; \, z=  z_v = (0,i\sinh  v,\cosh  v),\  \, v<0\}
\end{equation}
\end{proposition}
\noindent
\begin{proof}
It is sufficient to consider the case of
${\mathcal T}_\rightarrow $
and $h_\rightarrow$.
The action of the one-parameter subgroup $G_b$ on $h_\rightarrow$ generates the set
\begin{equation}
G_b\  h_\rightarrow = \{ z  \in X^{(c)}; \,
\label{Gbh}
z= (i\sinh v\sinh \alpha,\ i\sinh v\cosh \alpha,\  \cosh v),\  \, v>0\}
\end{equation}
Let now $z= x+iy$ be an arbitrary point in ${\mathcal T}_\rightarrow$; in view of
(\ref{Tell}), $x$ belongs to an hyperboloid of the form $x^2 = - \cosh^2 v $,
and therefore  (by transitivity) there exists a $g \in G$ such that
$gx = b_v = { \cosh v}\  b$.
Then the point $gy$ must be such that $[gy\cdot b_v]= [y\cdot x] =0$
with $(gy)^2=-\sinh^2 v$ and $\epsilon(b_v+igy)= -$,
which implies that $gz = b_v + i gy$ is of the form (\ref{Gbh});
therefore one has $gz \in G_b h_\rightarrow$ and $z \in G h_\rightarrow$.
\end{proof}

\vskip 0.3cm The tube ${\mathcal T}_\rightarrow$ is a tuboid over
$X$ in $X^{(c)}$ whose profile at the base point $b$ is the cone
$V_\rightarrow (b) = \{y\in {\mathbb R}^3; y_{1}> |y_{0}|,\
y_{2}=0 \}$; the latter is in fact obtained by taking the limit $v
\to 0$ from the set $ G_b h_\rightarrow$ into the tangent plane to
$X$ at $b$. The tube ${\mathcal T}_\leftarrow$ is similarly
described by replacing the condition $v>0$ by $v<0$ in the
previous analysis: it is a tuboid whose profile at the point $b$
is the cone $V_\leftarrow(b) = - V_\rightarrow(b)$.

\vskip 0.4cm
\noindent
{\bf A simple parametrization of  ${\mathcal T}_\rightarrow$ and ${\mathcal T}_\leftarrow$.}

\vskip 0.3cm We now give a parametrization of  ${\mathcal
T}_\rightarrow$ and ${\mathcal T}_\leftarrow$ which exhibits these
domains as being {\sl semi-tubes} in ${\mathbb C}^2$. Let us
parametrize $X^{(c)}$ as follows:
\begin{equation}
z= z[\theta,\Psi]\equiv  (\sinh  \Psi,\  \cosh \Psi \sin\theta,\  \cosh \Psi \cos\theta)
\label{param}
\end{equation}
with $(\theta = u + iv, \Psi = \psi + i \varphi) \in {{\mathbb
C}^2/(2\pi {\mathbb Z})^2}$. In this parametrization, the
translations $u\rightarrow u+a$ represent the rotations with axis
$Oz_0$, which leave the domains ${\mathcal T}_\rightarrow$ and
${\mathcal T}_\leftarrow$ invariant. One thus expects that the
latter are represented by semi-tubes in ${\mathbb C}^2$ bordered
by surfaces $v= v_\pm(\psi,\varphi) $ in $(\theta,\Psi)-$space. By
taking the imaginary parts in Eq.(\ref{param}),  one rewrites the
defining condition of ${\mathcal T}_{\chi}$ as follows:
\begin{equation}
y^2 = \sin^2 \varphi - \sinh ^2 v(\cosh ^2 \psi -\sin^2 \varphi)\ \   < 0,
\end{equation}
which implies the following representation for
${\mathcal T}_\rightarrow$ and
${\mathcal T}_\leftarrow$:
\begin{equation}
{\mathcal T}_\rightarrow
\,\,\,\,\,\,\, \tanh  v > \frac{|  \sin \varphi|}{\cosh\psi}
\label{Tright}
\end{equation}
\begin{equation}
{\mathcal T}_\leftarrow
\,\,\,\,\,\,\, \tanh  v < -\frac{|  \sin \varphi|}{\cosh\psi}
\label{Tleft}
\end{equation}

\vskip 0.4cm
\noindent
We shall now prove the following property which is the analogue of Proposition 3:
\begin{proposition}
The projection of the domain ${\mathcal T}_\rightarrow$ (or
${\mathcal T}_\leftarrow$) in the complex plane of the coordinate
$z_2 = -[z\cdot b]$ is the cut-plane $\Theta_{\chi} = {\mathbb C}
\setminus [-1,1].$ The image of the domain ${\mathcal
T}_\leftarrow \times {\mathcal T}_\rightarrow$ (or ${\mathcal
T}_\rightarrow \times {\mathcal T}_\leftarrow$) by the mapping
$(z,z') \rightarrow [z\cdot z']$ is the cut-plane $\Theta_{\chi}$.
\end{proposition}
\begin{proof}
a) In view of Eq. (\ref{param}), we have:
\begin{equation}
z_{2}=\cosh \left( \psi +i\varphi \right) \cos \left( u+iv\right)
\label{z}
\end{equation}
Eq. (\ref{Tright}) shows that ${\mathcal T}_\rightarrow$ contains
the set parametrized by $\{(\theta, \Psi);\ \theta = u+ iv;\ $  $
u \in {\mathbb R},\  v >0,\ \Psi = \psi + i \varphi =0 \}$, whose
image in the $z_2-$plane is (in view of (\ref{z})) exactly
$\Theta_{\chi}$.

Let us now show that the points exterior to $\Theta_{\chi}$, namely the points
$z_2 \in [-1,+1]$ cannot be the projections of points $z$ in
${\mathcal T}_\rightarrow$ or in ${\mathcal T}_\leftarrow$.
In fact, all points $z=(z_0,z_1,z_2) \in X^{(c)}$ with
$z_2^2<1$ belong to a complex hyperbola $h(z_2)$
with equation $ {\hat z}^2 \equiv z_0^2 - z_1^2 = -(1-z_2^2)   <0$.
All the complex points $z= (\hat z, z_2)$,
with $\hat z =(z_0,z_1) = \hat x + i\hat y$ in $h(z_2)$ satisfy the
conditions ${\hat x}^2 - {\hat y}^2 <0$ and $ [{\hat x}\cdot {\hat y}] =0$, which imply
${\hat x}^2 <0$ and ${\hat y}^2 \equiv y^2 >0$ and therefore $ z \notin {\mathcal T}_{\chi}$.
As for the points $z$ such that $z_2^2 =1$, they are such that ${\hat x}^2 = y^2 =0$ and
therefore also not in ${\mathcal T}_{\chi}$.

\noindent b) In view of Proposition 5, it is sufficient to check
that the image of the domain $\left\{ \left( z',z\right);z' \in
h_{\leftarrow },z \in {\mathcal T}_{\rightarrow }\right\}$ into
the plane of the variable $[z\cdot z']$ is $\Theta_{\chi}$. Let
therefore
\begin{equation}
z' = (0, -i \sinh v', \cosh v'),\ \  {\rm with}\ \  v' >0,
\end{equation}
while $z$ is parametrized as in Eq.(\ref{param}), the inequality (\ref{Tright}) being
satisfied.

Let $g_{v'}$ be the complex rotation with axis $Oz_0$ and angle $iv'$ whose effect is to
change $z'$ into $b$. The action of this
rotation on $z$ gives:
\begin{equation}
g_{v'}z =\left( \sinh \Psi,\ \cosh \Psi \sin \left( u+i\left(
v+v'\right) \right) \right) ,\  \cosh \Psi \cos \left( u+i\left( v+v'\right) \right)
\label{rot}
\end{equation}
and one thus has:
\begin{equation}
[b\cdot g_{v'}z] =[z'\cdot z] =\cosh \Psi \cos \left(
u+i\left( v+v'\right) \right) .
\end{equation}
Since $v'>0,$ it follows that:
\begin{equation}
\tanh \left( v+v'\right) >\tanh v>\frac{\left| \sin
\varphi \right| }{\cosh \psi},
\label{rota}
\end{equation}
which shows that $g_{v'}z \in {\mathcal T}_{\rightarrow }$ and in view of
a), that $[z\cdot z']\in \Theta_{\chi} .$ Of course all the points of $\Theta_{\chi} $ are
obtained in this image, since (in view of (\ref{rot}),(\ref{rota})), $g_{v'}z$ varies in the whole set
${\mathcal T}_{\rightarrow }$ when $z$ varies in the latter and $v'$ takes all positive values.
\end{proof}

\section{Cauchy-type representation in the tuboids ${\mathcal T}^+, {\mathcal T}^-,$
${\mathcal T}_\rightarrow$,
${\mathcal T}_\leftarrow$
and holomorphic decomposition on the one-sheeted hyperboloid}

\subsection{A global parametrization of the four tuboids}
It is convenient to use the following parametrization of $\hat X^{(c)} =
\{z \in X^{(c)}; \ z_0 + z_1 \ne 0\}$:

$$ z= z(\lambda, \mu): $$
\begin{equation}
z_0=\frac{1+\lambda\mu}
{\lambda - \mu},\ \
z_1=\frac{1- \lambda\mu}
{\lambda - \mu},\ \
z_2=\frac{\lambda+\mu }
{\lambda - \mu}\ \
\label{paragen}
\end{equation}
$${\rm with} \ \ (\lambda, \mu) \in {\mathbb C}^2 \setminus \delta,$$
$\delta$ being the diagonal $ (\lambda = \mu)$, which represents
points at infinity of $X^{(c)}$. The holomorphic compactification
${\mathbb S}_2 \times {\mathbb S}_2 $ of ${\mathbb C}^2 \setminus
\delta$ thus provides a corresponding compactification of
$X^{(c)}$ by an extension of the bijective mapping
(\ref{paragen}).

\vskip 0.3cm
The inversion formulae
\begin{equation}
\lambda = \frac{z_0-z_1}{z_2-1}=\frac{z_2+1}{z_0+z_1},\ \
\mu = \frac{z_0-z_1}{z_2+1}=\frac{z_2-1}{z_0+z_1}
\label{invparagen}
\end{equation}
exhibit the (real or complex) lines $\lambda = $cst and $\mu = $cst as the two systems of
(real or complex) linear generatrices of $X^{(c)}$.

\vskip 0.3cm In the space ${\mathbb C}^2$ of the variables
($\lambda, \mu$)  we introduce the four tubes whose imaginary
bases are the coordinate quadrants, denoted as follows in terms of
sign-valued functions $ \varepsilon_{\lambda}$ and
$\varepsilon_{\mu}$:
$$ \tau^{\varepsilon_{\lambda}, \varepsilon_{\mu}} = \{(\lambda,\mu) \in {\mathbb C}^2;
\ \varepsilon_{\lambda}\ {\rm Im}\lambda >0,\
\ \varepsilon_{\mu}\ {\rm Im}\mu >0,\}.$$
We shall then prove:

\begin{proposition}
Formulae (\ref{paragen}) and (\ref{invparagen}) define biholomorphic mappings from the
tubes $\tau^{-,+}$ and $\tau^{+,-}$
onto the respective Lorentz tuboids
${\mathcal T}^+$ and ${\mathcal T}^-$
and from the ``pierced'' tubes
$\tau^{+,+}\setminus \delta$ and $\tau^{-,-}\setminus \delta$
onto the respective chiral tuboids
${\mathcal T}_\leftarrow$ and
${\mathcal T}_\rightarrow$.
\end{proposition}

\begin{proof}
We first compute the bilinear form $[z\cdot \xi] = z_0\xi_0 - z_1\xi_1 -z_2\xi_2$ for $z \in X^{(c)}$ and
$\xi = \xi(\alpha) =(1,\cos {\alpha}, \sin {\alpha}) \in C^+,\ |\alpha| \leq \pi$. In view of (\ref{paragen}), one
obtains, for $|\alpha| \ne \pi$:
\begin{equation}
[z(\lambda,\mu)\cdot \xi(\alpha)] =
2 \cos^2 {\alpha/2}\  \frac{(\lambda- \tan {\alpha/2})(\mu - \tan {\alpha/2})}
{\lambda - \mu},
\label{zxi}
\end{equation}
or by putting
\begin{equation}
\lambda_{\alpha} = - \frac{1}{\lambda - \tan {\alpha/2}},\ \ \
\mu_{\alpha} = - \frac{1}{\mu - \tan {\alpha/2}} \label{invalpha}
\end{equation}
it follows
\begin{equation}
[z\cdot \xi(\alpha)] = 2 \cos^2 {\alpha/2}\  \frac{1}{\lambda_{\alpha} - \mu_{\alpha}}.
\label{paragenFH}
\end{equation}
\vskip 0.3cm
\noindent
a) Since for $|\alpha| \ne \pi$, the quantities ${\rm Im} \zeta $ and
${\rm Im} \zeta_{\alpha} $ have the same sign, it follows from Eq.(\ref{paragenFH}) that
the condition $(\lambda,\mu) \in \tau^{-,+}$
(resp. $(\lambda,\mu) \in \tau^{+,-}$) implies for $z=z(\lambda,\mu)$ the inequality
${\rm Im}([z\cdot\xi(\alpha)]) >0$
(resp. ${\rm Im}([z\cdot\xi(\alpha)]) <0$ ) for all values of $\alpha$ between $-\pi$ and $\pi$.
For $\alpha = \pm \pi$, one has $[z\cdot \xi(\pm\pi)] = z_0 + z_1 =
2 \ (\lambda - \mu)^{-1} $  and the same
implications still hold.
Therefore, in view of the characterization of ${\mathcal T}^{\pm}$ given in Proposition 1,
we have shown that the images of $\tau^{-,+}$ and $\tau^{+,-}$ by the mapping
$(\lambda,\mu) \to z(\lambda,\mu)$ are respectively contained in
${\mathcal T}^{+}$
and ${\mathcal T}^{-}$.
\vskip 0.3cm
\noindent
b) We shall now show that for any point $(\lambda,\mu)$ in $\tau^{+,+} \setminus \delta$, there exist two
distinct real numbers
$t_+ = \tan {\alpha_+/2}$ and
$t_- = \tan {\alpha_-/2}$
such that the corresponding complex numbers (defined by (\ref{invalpha}))
$\lambda_{\alpha_{\pm}}$,
$\mu_{\alpha_{\pm}}$
satisfy the equalities
\begin{equation}
{\rm Im} (\lambda_{\alpha_+}- \mu_{\alpha_+}) =\
{\rm Im} (\lambda_{\alpha_-}- \mu_{\alpha_-}) = 0,
\label{sameimag}
\end{equation}
so that (in view of (\ref{paragenFH})) the corresponding image $z=z(\lambda,\mu)$ satisfies the
equalities:
\begin{equation}
{\rm Im }\left([z\cdot\xi(\alpha_+)]\right)
= {\rm Im }\left([z\cdot\xi(\alpha_- )]\right) =0
\label{2tgtplanes}
\end{equation}
In fact, being given $\lambda $ and $\mu$ distinct in the upper half-plane,
the numbers $t_+$ and $t_-$ are determined by the following geometrical procedure:
there are two circles $\gamma_+$ and $\gamma_-$
which contain the points $\lambda$ and $\mu$ and are tangent to the real axis
at the respective points $t_+$ and $t_-$. These points lead to the desired condition for the following
reason: the inversion $\zeta - t_+ {\to} - (\zeta -t_+)^{-1}$ transforms the circle $\gamma_+$ into
a straight line parallel to the real axis ; therefore the points
$\lambda_{\alpha_{+}}$ and
$\mu_{\alpha_{+}}$ are on this line and thereby satisfy Eq.(\ref{sameimag}) (the same holds for $t_-$).
Note that if ${\rm Im} (\lambda - \mu) =0$, one of the points, say $t_+$, is rejected at infinity:
this is the case when $\alpha_+ = \pm \pi$, Eq. (\ref{2tgtplanes}) being still valid.

Since Eq.(\ref{2tgtplanes}) holds, it results from the characterization of ${\mathcal T}_{\chi}$ given
in Proposition 4
that the image of $\tau^{+,+}\setminus \delta$ by the mapping
$(\lambda,\mu) \to z(\lambda,\mu)$ is contained in
${\mathcal T}_{\chi}$. It remains to check that it is contained in the chiral component
${\mathcal T}_{\leftarrow}$ of the latter, defined by (\ref{chirality}); in fact, one has:
$\epsilon(z) = (i/2) ( z_1 \overline z_2 - z_2 \overline z_1)
= \left[ {\rm Im}\lambda (1 + |\mu|^2)
 +{\rm Im}\mu (1 + |\lambda|^2)\right]\times |\lambda - \mu|^{-2}$,
 which is positive for all $(\lambda, \mu)$ in $\tau^{+,+}\setminus \delta.$
Similarly, the image of $\tau^{-,-}\setminus \delta$ is shown to be contained in
${\mathcal T}_{\rightarrow}$.

Finally, in view of (\ref{invparagen}), the set of points
$(\lambda,\mu)$ in ${\mathbb C}^2$ such that either $\lambda $ or
$\mu$ is real represent complex points $z= x=iy$ of $X^{(c)}$
which belong to the complexified linear generatrices of $X$: all
these points are such that $y^2 =0$, namely they belong to the set
${\mathcal T}_0 \setminus \{z \in X^{(c)};\ z_0 + z_1 =0\} $.

So we have established that the biholomorphic mapping defined by
Eqs (\ref{paragen}), (\ref{invparagen}), maps respectively each
set of the partition of ${\mathbb C}^2_{(\lambda,\mu)}\setminus
\delta$ composed of the tubes $\tau^{-,+}, \tau^{+,-},$ $
\tau^{-,-}\setminus\delta,\  \tau^{+,+}\setminus \delta$ and of
their interfaces into a corresponding set of the partition of
$X^{(c)}\setminus \{z \in X^{(c)};\ z_0 + z_1 =0\} $ composed of
the tuboids ${\mathcal T}^+, {\mathcal T}^-,$ ${\mathcal
T}_\rightarrow,$ ${\mathcal T}_\leftarrow$ and of ${\mathcal T}_0
\setminus \{z \in X^{(c)};\ z_0 + z_1 =0\} $. Since the whole
space ${\mathbb C}^2_{(\lambda,\mu)}\setminus \delta$ is
biholomorphically mapped {\sl onto} the whole manifold
$X^{(c)}\setminus
 \{z \in X^{(c)};\ z_0 + z_1 =0\} $, it follows that the mapping is a {\sl bijection} for all corresponding
sets of the previous partitions, which ends the proof of the Proposition.
\end{proof}

\begin{remark}
The previous proposition shows that, while the tuboids
${\mathcal T}^+$ and ${\mathcal T}^-$ are simply-connected domains, the other two tuboids
${\mathcal T}_\rightarrow$ and
${\mathcal T}_\leftarrow$ admit a nontrivial homotopy generator, corresponding to a loop around
the set $\delta$ in their representations by the domains
$ \tau^{-,-}\setminus\delta,\  \tau^{+,+}\setminus \delta$. On the manifold $X^{(c)}$, a
typical representative of such a generator is the ``circle'' of the plane $z_0 =0$
parametrized by $z_1 = \sin (u+iv_0),\ z_2 = \cos (u +iv_0)$; in fact,
in view of Proposition 5 (or of the representation (\ref{param}),
(\ref{Tright}),
(\ref{Tleft}) with $\Psi = 0$),
such a circle is contained in
${\mathcal T}_\rightarrow$ or in
${\mathcal T}_\leftarrow$ according to whether the constant $v_0$ is positive or negative
(note that when $z$ describes this circle,
the variable ${\lambda - \mu} = 2/ z_1$ describes a loop around the origin).
\end{remark}

\subsection{Holomorphic functions in the tuboids and integral representations}

We shall make use of the
following relations which are direct consequences of (\ref{leray}) and (\ref{paragen}):
\begin{equation}
{\rm d}\sigma(z(\lambda,\mu)) = \frac{ {\rm d}(z_0 + z_1) {\rm d}z_2}{2\ |z_0 + z_1|}
= \frac{{\rm d}\lambda {\rm d} \mu}{(\lambda - \mu)^2}
\label{measure}
\end{equation}
and
\begin{equation}
[z(\lambda,\mu)- z(\lambda',\mu')]^2
= -\frac{4(\lambda-\lambda')(\mu-\mu') }{(\lambda - \mu)(\lambda'-\mu')}
\label{cauchy}
\end{equation}
In view of (\ref{measure}), the Hilbert space ${\mathcal H}(X)$ of
functions $f(x)$ on $X$ which are square-integrable with respect
to the measure ${\rm d}\sigma(x)$ can be represented by the space
$\hat{\mathcal H}$ of functions $\hat f(\lambda,\mu) =
f(x(\lambda,\mu))$ in $L^2\left( {\mathbb R}^2,\ \frac{{\rm
d}\lambda {\rm d}\mu}{(\lambda - \mu)^2}\right)$.

\vskip 0.4cm For each of the four tuboids ${\mathcal T}^+,
{\mathcal T}^-,$ ${\mathcal T}_\rightarrow,$ and ${\mathcal
T}_\leftarrow$ of $X^{(c)}$, we now denote respectively
$H^2\left({\mathcal T}^+\right)$, $H^2\left({\mathcal
T}^-\right)$, $H^2 \left({\mathcal T}_\rightarrow\right)$  and
$H^2 \left({\mathcal T}_\leftarrow\right)$ the space of functions
$F(z)$, which enjoy the following properties:

a) $F$ is holomorphic in the tuboid considered,

b) $F$ admits a boundary value $f(x)$ on $X$ from this tuboid, which belongs to ${\mathcal H}(X)$,

c) $F$ is ``sufficiently regular at infinity in its domain'' in the following sense:
the inverse image $\hat F(\lambda,\mu) = F(z(\lambda,\mu))$ of $F$
is such that
$(\lambda - \mu)^{-1}
\hat F(\lambda, \mu)$
belongs to the Hardy space
$H^2(\tau^{\epsilon_{\lambda}, \epsilon_{\mu}})$ of the corresponding tube
$\tau^{\epsilon_{\lambda}, \epsilon_{\mu}}$
(with $\epsilon_{\lambda}, \epsilon_{\mu} = \pm 1$
given by
the prescription of Proposition 7).

\vskip 0.4cm
Under these assumptions,
we can therefore write the following Cauchy integral representation
in two variables for the
function
$(\lambda - \mu)^{-1}
\hat F(\lambda, \mu)$:
\begin{equation}
\frac{\hat F(\lambda,\mu)}{\lambda - \mu} = -{\epsilon_{\lambda}
\epsilon_{\mu}} \frac{1}{4\pi^2} \int_{{\mathbb R}^2} \frac {\hat
f(\lambda',\mu')} {\lambda'-\mu'} \frac{{\rm d} \lambda' {\rm d}
\mu'} {(\lambda'-\lambda)(\mu'-\mu)}. \label{cauchyrep}
\end{equation}
By rewriting the latter as follows:
\begin{equation}
\hat F(\lambda,\mu) = -{\epsilon_{\lambda}  \epsilon_{\mu}}
\frac{1}{\pi^2} \int_{{\mathbb R}^2} \hat f(\lambda',\mu') \frac
{(\lambda - \mu)(\lambda'-\mu')}{4(\lambda-\lambda')(\mu-\mu') } \
\frac{{\rm d}\lambda' {\rm d} \mu'} {(\lambda' - \mu')^2},
 \label{klmnopqr}
\end{equation}
we can then take Eqs (\ref{measure}) and (\ref{cauchy}) into account and obtain
the following Cauchy-type representation for the functions of the previous
spaces, holomorphic in
either one of the four tuboids
${\mathcal T}^+, {\mathcal T}^-,$
${\mathcal T}_\rightarrow$ and
${\mathcal T}_\leftarrow$ of $X^{(c)}$:
\begin{equation}
F(z)={\mp 1 \over \pi^ 2} \int_X {f(x) \over (x-z)^2}\  {\rm d} \sigma (x)
\label{Cauchy}
\end{equation}
In the r.h.s. of the latter, the sign -- corresponds to the case of the tuboids
${\mathcal T}^+$ and $ {\mathcal T}^-,$
while the sign + corresponds to the case of the tuboids
${\mathcal T}_\rightarrow$ and
${\mathcal T}_\leftarrow$.

\begin{remark}
One checks that for every point $x \in X$ the singular set of the Cauchy kernel of
(\ref{Cauchy}), namely $\{z;\ (x-z)^2 =0\}$, does not intersect the previous four tuboids.
In fact, this singular set coincides with the intersection of $X^{(c)}$ with its
analytic tangent plane at $x$ (whose  equation is $[(z-x)\cdot x] = [z\cdot x] +1 = 0$);
it is therefore composed of the two linear generatrices of $X$ containing $x$,
whose complex points are all in ${\mathcal T}_0$.
\end{remark}

\subsection{Invariance properties of the spaces ${\mathcal H}(X)$ and
$H^2\left({\mathcal T}^+\right),
H^2\left({\mathcal T}^-\right),$
$H^2 \left({\mathcal T}_\rightarrow\right),$
$H^2 \left({\mathcal T}_\leftarrow\right)$}

The invariance of the measure ${\rm d}\sigma(x)$ and of the Cauchy
kernel $\left[(x-z)^2\right]^{-1}$ under the group $G$ is
equivalently represented by the conformal invariance of the
corresponding quantities expressed in terms of the previous set of
variables $(\lambda,\mu)$. Apart from the invariance under the
translations and dilatations, which is trivial, we shall stress
the invariance under the homographic transformations
$(\lambda,\mu) \to (\lambda_{\alpha},\mu_{\alpha})$ defined by Eq.
(\ref{invalpha}) and whose inverses are given by the formulae
$$ \lambda = \tan {\alpha/2} - 1/\lambda_{\alpha},\ \mu = \tan {\alpha/2} - 1/\mu_{\alpha}$$
In fact, one checks that one has for all values of $\alpha$:
\begin{equation}
{\rm d}\sigma(x(\lambda,\mu)) =
\frac{{\rm d}\lambda {\rm d} \mu}{(\lambda - \mu)^2}
= \frac{{\rm d}\lambda_{\alpha} {\rm d} \mu_{\alpha}}{(\lambda_{\alpha} - \mu_{\alpha})^2}
\label{invmeasure}
\end{equation}
and
\begin{equation}
[z(\lambda,\mu)- z(\lambda',\mu')]^2
= -\frac{4(\lambda-\lambda')(\mu-\mu') }{(\lambda - \mu)(\lambda'-\mu')}
= -\frac{4(\lambda_{\alpha}-\lambda'_{\alpha})(\mu_{\alpha}-\mu'_{\alpha}) }
{(\lambda_{\alpha} - \mu_{\alpha})(\lambda'_{\alpha}-\mu'_{\alpha})}
\label{invcauchy}
\end{equation}
In view of (\ref{invmeasure}), the Hilbert space ${\mathcal H}(X)$
can then be represented, for all $\alpha's$, by the space of
functions $\hat f_{\alpha}(\lambda_{\alpha},\mu_{\alpha}) =
f(x(\tan {\alpha/2} -1/\lambda_{\alpha},\tan{ \alpha/2} -
1/\mu_{\alpha}))$ in $L^2\left( {\mathbb R}^2,\ \frac{{\rm
d}\lambda_{\alpha} {\rm d}\mu_{\alpha}}{(\lambda_{\alpha} -
\mu_{\alpha})^2}\right)$.

Similarly each function $F(z)$ in either one of the spaces
$H^2\left({\mathcal T}^+\right),
H^2\left({\mathcal T}^-\right),$
$H^2 \left({\mathcal T}_\rightarrow\right),$
or $H^2 \left({\mathcal T}_\leftarrow\right)$
is represented for all values of $\alpha$ by a holomorphic function
$\hat F_{\alpha}(\lambda_{\alpha}, \mu_{\alpha}) =
F(z(\tan {\alpha/2} -1/\lambda_{\alpha},\tan{ \alpha/2} - 1/\mu_{\alpha}))$
whose domain is the corresponding tube $\tau^{\epsilon_{\lambda},\epsilon_{\mu}}$.
Moreover, in view of (\ref{invmeasure}) and (\ref{invcauchy}),
the formula (\ref{cauchyrep}) is seen to be invariant under the transformation
$(\lambda,\mu) \to (\lambda_{\alpha},\mu_{\alpha})$,
which shows that each function
$(\lambda_{\alpha} - \mu_{\alpha})^{-1}
\hat F_{\alpha}(\lambda_{\alpha},\mu_{\alpha})$
belongs to the same Hardy space
$H^2\left(\tau^{\epsilon_{\lambda},\epsilon_{\mu}}\right)$.

{\sl This invariance allows us legitimately to call the spaces
$H^2\left({\mathcal T}^+\right),$
$H^2\left({\mathcal T}^-\right),$
$H^2 \left({\mathcal T}_\rightarrow\right)$
and $H^2 \left({\mathcal T}_\leftarrow\right)$
Hardy spaces of the corresponding tuboids of $X^{(c)}$.}

\subsection{ Decomposition in Hardy spaces of the four tuboids}

Every function $\hat f(\lambda, \mu)$ in $\hat {\mathcal H}$ admits a
decomposition of the form
\begin{equation}
\hat f = \hat f^{+,+} + \hat f^{-,-} +\hat f^{+,-} + \hat f^{-,+},
\label{dec}
\end{equation}
where each function $\hat f^{\epsilon_{\lambda}, \epsilon_{\mu}}$ is the boundary value
of a holomorphic function
$\hat F^{\epsilon_{\lambda}, \epsilon_{\mu}}$, such that
$(\lambda - \mu)^{-1}
\hat F(\lambda,\mu)$
belongs to the Hardy space
$H^2(\tau^{\epsilon_{\lambda}, \epsilon_{\mu}})$ of the corresponding tube
$\tau^{\epsilon_{\lambda}, \epsilon_{\mu}}$.
Each function
$\hat F^{\epsilon_{\lambda}, \epsilon_{\mu}}$ satisfies the Cauchy integral representation
(\ref{cauchyrep})
and is also directly defined in terms of $\hat f$ by the same Cauchy integral
in which $\hat f$ is substituted to
$\hat f^{\epsilon_{\lambda}, \epsilon_{\mu}}$.
This standard result is most simply obtained by considering the
four holomorphic functions
$(\lambda - \mu)^{-1}
\hat F^{\epsilon_{\lambda}, \epsilon_{\mu}}(\lambda,\mu) $
as the
inverse Fourier-Laplace transforms
of the Fourier transform of
$(\lambda - \mu)^{-1}
\hat f(\lambda,\mu)$,
chopped with the characteristic
functions of the four quadrants of Fourier coordinates.

\vskip 0.4cm
By applying the results of subsections 3.2 and 3.3, and in particular formula (\ref{Cauchy}),
we then immediately obtain the following

\vskip 0.4cm
\noindent
{\bf Theorem 1}
\ {\sl Every function $f(x)$ in ${\mathcal H}(X)$ admits a decomposition of the form
\begin{equation}
f = f^{+} + f^{-} +f_{\rightarrow} + f_{\leftarrow} = {\sum}_{tub}\  f_{(tub)} ,
\label{Dec}
\end{equation}
in which each of the four components $f_{(tub)}(x)$ is the boundary value in ${\mathcal H}(X)$
of a holomorphic function $F_{(tub)}(z)$ belonging to the corresponding Hardy space
$H^2\left({\mathcal T}^+\right),$
$H^2\left({\mathcal T}^-\right),$
$H^2 \left({\mathcal T}_\rightarrow\right)$
and $H^2 \left({\mathcal T}_\leftarrow\right)$.
Moreover, each function
$F_{(tub)}(z)$ is given in its tuboid by the following
integral representations (expressed either in terms of $f$ or of its own boundary
value $f_{(tub)}$):
\begin{equation}
F_{(tub)} (z)={\epsilon_{(tub)}}\ {1 \over \pi^ 2} \int_X {f(x) \over (x-z)^2}\  {\rm d} \sigma (x)
={\epsilon_{(tub)}}\ {1 \over \pi^ 2} \int_X {f_{(tub)}(x) \over (x-z)^2}\  {\rm d} \sigma (x)
\label{Cauchydec},
\end{equation}
in which the sign function
${\epsilon_{(tub)}}$ takes the value -- for the tuboids
${\mathcal T}^+, {\mathcal T}^-,$ and +  for the tuboids
${\mathcal T}_\rightarrow$
and ${\mathcal T}_\leftarrow$.}

\vskip 0.5cm
We end this section by introducing dense subspaces of the previous spaces
which will be of current use in the study of the Fourier-Helgason transformation.

\vskip 0.5cm
\noindent
{\bf Definition 1}
{\sl We call
${\mathcal H}_{(reg)}(X)$
and $H^2_{(reg)}\left({\rm tub}\right),$ where ${\rm tub}$ stands for either one of the four tuboids
${\mathcal T}^+,$ $ {\mathcal T}^-,$
${\mathcal T}_\rightarrow,$
${\mathcal T}_\leftarrow,$
the respective subspaces of
${\mathcal H}(X)$ and
$H^2\left({\rm tub}\right) $
which are represented in the variables $(\lambda,\mu)$ by the set of all functions
$\hat f(\lambda,\mu)$ in the corresponding space $\hat{\mathcal H}$ or $H^2(\tau^{\epsilon_{\lambda},
\epsilon_{\mu}})$
satisfying the following boundedness property:

\noindent $(\lambda - \mu)^{-1} \hat f(\lambda,\mu)$ (resp.
$(\lambda - \mu)^{-1} \hat F(\lambda,\mu)$) admits a uniform bound
on ${\mathbb R}^2$ (resp. in the closure of the tube
$\tau^{\epsilon_{\lambda}, \epsilon_{\mu}}$) of the form $ {\rm
cst} \times (1+|\lambda|)^{-1}(1+|\mu|)^{-1}$, where ${\rm cst}$
denotes an arbitrary constant.}

\section{The Lorentzian Fourier-Helgason transformation}

\subsection{Definition and properties of the transforms $\tilde f_+, \tilde f_-$}

For all functions $f(x)$ in ${\mathcal H}_{(reg)}(X)$
we introduce the following pair of transforms
\begin{equation}
\tilde{f}_{\pm}(\xi,s) = \int_{X}
[x_\pm\cdot \xi]^{s} f(x)\ {\rm d}\sigma(x)
\label{fourier1}
\end{equation}
in which $s$ is a complex parameter with appropriate range, $\xi$
varies on the cone $C^+$ and the kernels $[x_+\cdot \xi]^{s}$,
$[x_-\cdot \xi]^{s}$ are defined, for each $\xi \in C^+$, as the
boundary values of the holomorphic function $[z\cdot \xi]^s =
[(x+iy)\cdot \xi]^s$ when $z$ tends to the reals from the
respective tuboids ${\mathcal T}^+$ and ${\mathcal T}^-$ of
$X^{(c)}$.  In fact, in view of Proposition 1, for each $\xi \in
C^+$ the function $[z\cdot \xi]^s$ is holomorphic in the union  of
${\mathcal T}^+$ and ${\mathcal T}^-,$ since $[z \cdot\xi]$ takes
its values correspondingly in the upper and lower half-planes.
However, this statement necessitates the following specification:
putting $[z\cdot \xi] =t$, the holomorphic function $t^s$ is
always considered in its distinguished sheet over ${\mathbb C}
\setminus ]-\infty,-1]$, namely, as being positive on ${\mathbb
R}^+$. {\sl Denoting by $t_+^s$ and $t_-^s$ the boundary values on
${\mathbb R}$ of the holomorphic function $t^s$ respectively from
the upper and lower half-planes,} one then has the identity:
\begin{equation}
[x_\pm\cdot \xi]^{s} =  [x\cdot\xi]_{\pm}^s
= Y([x\cdot\xi])|[x\cdot\xi]|^s +
e^{\pm i\pi s} Y(-[x\cdot\xi])|[x\cdot\xi]|^s,
\label{detail}
\end{equation}
in which $Y(t)$ is the Heaviside function ($Y(t) =1$ for $t>0$ and $=0$ for $t <0$).
Note that, below, one will only deal with the case ${\rm Re}s > -1$ so that the previous
equality (\ref{detail}) will always hold in the sense of functions in $L^1$. (A
distribution-like treatment would be necessary only in higher dimensions, as considered in \cite{[BM]}).

\vskip 0.3cm
The transforms $\tilde f_{\pm}$ of $f$ satisfy the homogeneity property
$\tilde f_{\pm}(r\xi,s) =
r^s \tilde f_{\pm}(\xi,s) $; by taking Eqs (\ref{zxi}) and (\ref{measure}) into account
for each vector $\xi = \xi(\alpha)=(1,\cos\alpha, \sin\alpha) \in C^+$,
Eq. (\ref{fourier1}) can be rewritten as follows in terms of the parametrization
(\ref{paragen}) of $X$:
\begin{equation}
\tilde{f}_{\pm}(\xi(\alpha),s) =  2^s  \int_{{\mathbb R}^2}
\left[\frac{\hat f (\lambda,\mu)}{\lambda - \mu}\right]
\frac{[(\cos \alpha/2) \lambda - (\sin \alpha/2)]_{\mp}^s [(\cos
\alpha/2) \mu-(\sin \alpha/2)]_{\pm}^s} {(\lambda -
\mu)_{\mp}^{s+1}} {\rm d}\lambda {\rm d}\mu. \label{fourier2}
\end{equation}
or equivalently (in view of Eqs (\ref{invalpha}),(\ref{paragenFH}) and (\ref{invmeasure})):
\begin{equation}
\tilde{f}_{\pm}(\xi(\alpha),s) =  (2 \cos^2 \alpha /2)^s
\int_{{\mathbb R}^2} \left[\frac{\hat f_{\alpha }
(\lambda_{\alpha},\mu_{\alpha})}{\lambda_{\alpha} -
\mu_{\alpha}}\right] \frac{{\rm d}\lambda_{\alpha} {\rm
d}\mu_{\alpha}} {(\lambda_{\alpha} - \mu_{\alpha})_{\mp}^{s+1}}
\label{fourier3}
\end{equation}
This allows us to state the following
\begin{proposition}
a) For every function $f$ in ${\mathcal H}_{(reg)}(X)$, the corresponding transforms
$\tilde{{f}}_{+}(\xi,s) $ and $\tilde{{f}}_{-}(\xi,s) $ are well-defined, continuous and homogeneous
of degree $s$ with respect to $\xi$ in $C^+$,
and holomorphic with respect
to $s$ in the strip
$-1 < {\rm Re} s <0. $

b) For every function $ f $ in the Hardy space $H^2_{(reg)}({\mathcal T}^+)$
(resp. $H^2_{(reg)}({\mathcal T}^{-}) $),
the corresponding transform
$\tilde{{f}}_{+}(\xi,s) $  (resp. $\tilde{{f}}_{-}(\xi,s)$ )  vanishes.
\end{proposition}
\begin{proof}

a) The following majorization of the r.h.s. of Eq. (\ref{fourier2}) results from
the uniform bound on $\hat f$
(postulated according to Definition 1)
together with the fact that $|(\lambda_{\alpha} - \mu_{\alpha})_{\mp} ^{-s} | \leq
{\rm max}(e^{\mp\pi\  {\rm Im}s},\ 1)  \times
|\lambda_{\alpha} - \mu_{\alpha}| ^{-{\rm Re}s}  $:
\[  |\tilde{f}_{\pm}(\xi(\alpha),s)|
\ \ \leq  {\rm Cst} \ {\rm max}(e^{\mp\pi\ {\rm Im}s},\ 1) \times \cdots \]
\begin{equation}
\cdots \int_{{\mathbb R}^2} \frac{|(\cos \alpha/2) \lambda- (\sin
\alpha/2)|^{{\rm Re}s}}{1+ |\lambda|} \frac{|(\cos \alpha/2) \mu
-(\sin \alpha/2)|^{{\rm Re} s}}{1+ |\mu|} \frac{{\rm d}\lambda
{\rm d}\mu} {|\lambda - \mu|^{{\rm Re}s+1}}. \label{majfourier2}
\end{equation}
The uniform convergence and boundedness of the latter integral
in all the intervals  $-1+ \eta \leq  {\rm Re}s \leq -\eta\ $ (with $\ \eta > 0\  $) gives the result.

b) Assuming that $ f $ belongs to $H^2_{(reg)}({\mathcal T}^+)$,
namely that $\hat f(\lambda,\mu)$ is the boundary value of a
holomorphic function $\hat F$ bounded by ${\rm cst}\times
(1+|\lambda|)^{-1}(1+|\mu|)^{-1}$ in $\tau^{-,+}$, we can distort
the integration cycle of (\ref{fourier2}) into ${\mathbb R}^2 + i
(-a,a)$, $a$ being any positive number, {\sl provided one has
chosen the prescription $\lambda_-, \mu_+ $ (of $\tilde f_+$) in
the integrand of (\ref{fourier2}).} The corresponding integral
being then independent of $a$, we see by taking the limit $a \to
\infty$ that this integral has to be equal to zero for all values
of $s$ in the strip $-1 < {\rm Re} s <0 $ where it is defined.
\end{proof}

\vskip 0.5cm
\noindent
{\bf Definition 2}\
{\sl Given a function $f$ in ${\mathcal H}_{(reg)}(X)$,
we define its {\rm Lorentzian Fourier-Helgason (FH) transforms}  as
the restrictions of $\tilde f_{\pm}$ to the symmetry axis $s = -1/2 -i \nu$ of
their analyticity domain in $s$, namely
the following pair of functions on the cone $C^+$, homogeneous of degree $-1/2 - i\nu$:
\begin{equation}
\tilde{f}_{\pm,\nu}(\xi ) =
\tilde{f}_{\pm}(\xi,-\frac{1}{2}-i\nu) =
\int_X [x_{\pm}\cdot \xi]^{-\frac{1}{2}-i\nu}
f(x)\ {\rm d}\sigma(x)
\label{Fourier}
\end{equation}
In view of proposition 8b),
for every function $ f \in H^2_{(reg)}({\mathcal T}^+)$
(resp. $H^2_{(reg)}({\mathcal T}^{-}) $), there is a {\rm unique} Lorentzian FH-transform which is
$\tilde{f}_{-,\nu}(\xi )$
(resp. $\tilde{f}_{+,\nu}(\xi )$).}

\vskip 0.3cm
\noindent
Note that in view of (\ref{detail}), the r.h.s. of Eq. (\ref{Fourier})
can be read more explicitly as follows:
\begin{equation}
\int_{\{x \in X; [x\cdot\xi] > 0\}}
|[x\cdot \xi]|^{-\frac{1}{2}-i\nu}
f(x)\ {\rm d}\sigma(x)
\mp i  \int_{\{x \in X; [x\cdot\xi] < 0\}}
e^{\pm \pi\nu}|[x\cdot \xi]|^{-\frac{1}{2}-i\nu}
f(x)\ {\rm d}\sigma(x).
\label{detailfourier}
\end{equation}
and that one has, in view of (\ref{majfourier2}), the following
\begin{proposition}
The Lorentzian FH-transforms $\tilde f_{\pm, \nu}$
of any function $f$ belonging to  ${\mathcal H}_{(reg)}(X)$ satisfy uniform bounds of the form:
\begin{equation}
|\tilde{f}_{\pm, \nu}(\xi)| \ \ \leq  {\rm Cst} \ \xi_0^{-1/2}\
{\rm max}(e^{\pm\pi \nu},\ 1) \label{majFH}
\end{equation}
\end{proposition}

\subsection{Inversion of the transformation}
Let $[i_{\Xi} \omega](\xi)$
be the one-form on $C^+$ obtained by contracting the vector field
$\Xi(\xi) = (\xi_0,\xi_1, \xi_2)$
with the $G-$invariant volume form
$\omega(\xi) = \left.\frac{{\rm d}\xi_0 \wedge {\rm d}\xi_1 \wedge {\rm d}\xi_2}
{{\rm d}(\xi^2)}\right|_{C^+}  $
and let ${\rm d}\mu_\gamma$ be the measure obtained by
restricting this one-form
to any given loop $\gamma$ on $C^+$
homotopic to the circle
$\gamma_0 = \{\xi \in C^+; \xi = \xi(\alpha) = (1,\cos \alpha,\sin \alpha), \  -\pi \leq 0 \leq \pi \}.$
One checks in particular that
${\rm d}\mu_{\gamma_0}
= {\rm d}\alpha /2. $
With Euler's identity, one checks the following property, of current use below:
\begin{proposition}
For every function $a(\xi)$ on $C^+$ homogeneous of degree $-1$, the one-form
$a(\xi)
[i_{\Xi} \omega](\xi)$ is closed.
\end{proposition}

We now wish to show:

\vskip 0.5cm
\noindent
{\bf Theorem 2}
{\sl Let $f(x)$ belong to
$H^2_{(reg)}\left({\mathcal T}^-\right)$
(resp. $H^2_{(reg)}\left({\mathcal T}^+\right)$),
and let $\tilde{f}_{+,\nu}$
(resp. $\tilde{f}_{-,\nu}$)
be its FH-transform.
Then the holomorphic function $F(z)$ in ${\mathcal T}^-$
(resp. ${\mathcal T}^+$) whose boundary value is $f$ is given in terms of
$\tilde{f}_{+,\nu}$
(resp. $\tilde{f}_{-,\nu}$)
by the following formula:
\begin{equation}
F(z) =  \frac{1}{2\pi^2} \int_0^{\infty}
\frac{\nu \tanh \pi \nu}{e^{\pm\pi\nu} \cosh \pi \nu} {\rm d}\nu
\int^{ }_ \gamma
[z\cdot \xi]^{-{1 \over 2}+i\nu} \tilde{f}_{\pm,\nu}
(\xi)\
{\rm d}\mu_ \gamma (\xi)
\label{invfourier}
\end{equation}}

One first checks that the double integral at the r.h.s. of
formula (\ref{invfourier}) converges
for all $z$ in
${\mathcal T}^-$
(resp. ${\mathcal T}^+$): in fact,
in view of Proposition 1, the condition $z\in
{\mathcal T}^-$
(resp. ${\mathcal T}^+$) implies that
$\left| [z\cdot \xi]^{i\nu}\right| \leq e^{-a \nu}$ for some $a = a(z)$ such that
$-\pi <  a < 0$  (resp. $0 < a < \pi$).
This implies, in view of Proposition 9, that the integrand at the r.h.s. of (\ref{invfourier})
is uniformly bounded by ${\rm cst} \ e^{-(\pi + a) \nu}$  (resp. ${\rm cst}\ e^{-a \nu}$). The
expression at the r.h.s. of Eq. (\ref{invfourier}) is therefore a holomorphic function of $z$
in ${\mathcal T}^-$
(resp. ${\mathcal T}^+$).
Moreover, due to the homogeneity
of degree $-\frac{1}{2} - i \nu$ in $\xi$ of $\tilde f_{\pm, \nu}$,
this holomorphic function is independent of the choice of the cycle
$\gamma$, as a consequence of Proposition 10 and Stokes theorem,

In order to prove Theorem 2,
we shall make use of the
Cauchy-type representation (\ref{Cauchy}) established above
together with the following expression of the Cauchy kernel on $X^{(c)}$
which will be proved in our next subsection:
\begin{proposition}
The Cauchy kernel on $X^{(c)}$ is given by the following double integral:
\begin{equation}
\frac{1}{(z'-z)^2} = -  \frac{1}{2}
\int_{0}^{\infty}
\frac{\nu \tanh \pi \nu}{e^{\pm \pi\nu} \cosh \pi \nu} {\rm d}\nu
\int^{ }_ \gamma
[z\cdot \xi]^{-{1 \over 2}+i\nu}
[\xi\cdot z']^{-{1 \over 2}-i\nu}
{\rm d}\mu_ \gamma (\xi)
\label{expcauchy}
\end{equation}
which is absolutely convergent for $(z,z')$ in ${\mathcal T}^- \times {\mathcal T}^+$
and  in ${\mathcal T}^+ \times {\mathcal T}^-$. This formula remains meaningful when one of the points,
e.g. $z'$, is taken real, provided the appropriate limit is taken, namely respectively
$z' =  x_+$ or $z' = x_-$, in the r.h.s. of (\ref{expcauchy}).
\end{proposition}

If we plug the expression (\ref{expcauchy}) of $\left[(x-z)^2\right]^{-1}$ into
the r.h.s. of (\ref{Cauchy}), invert the integrals over $x$ and over $(\nu,\xi)$ in this
absolutely convergent integral and take into account the defining formula
(\ref{Fourier}) of $\tilde f_{\pm,\nu}(\xi)$, we readily obtain formula (\ref{invfourier})
for both types of configurations $z \in {\mathcal T}^-, z' = x_+$ and
$z \in {\mathcal T}^+, z' = x_-$
(corresponding to the cases $f \in H^2_{(reg)}\left({\mathcal T}^-\right)$
and $f \in H^2_{(reg)}\left({\mathcal T}^+\right)$).
So, proving Theorem 2  amounts to proving Proposition 11.

\subsection{Lorentzian invariant perikernels: a new representation for the first-kind Legendre functions and
the Cauchy kernel on $X^{(c)}$}

An important class of holomorphic kernels on the ($d-$dimensional) complexified
hyperboloid has been introduced in (\cite{[BV]}) and
(\cite{[B]}) under the name of ``perikernels''.
\footnote
{These kernels arise in the context of quantum field theory on
a $d-$dimensional one-sheeted hyperboloid $(d \geq 2),$
interpreted as a $d-$dimensional de Sitter spacetime manifold.
In \cite{[BM]}
we have given a complete study of these kernels ${\mathcal W}(x_1,x_2)$ and characterized them
as the ``two-point functions'' of  Wightman quantum field theories on
the corresponding de Sitter spacetime.}
Since it plays a basic role in our approach
to the FH-transformation, we now recall this notion in the two-dimensional case presently
considered.

\vskip5pt
A perikernel is a holomorphic function ${\rm W}(z,z')$ defined in the ``cut-domain''
$\Delta = X^{(c)}\times X^{(c)} \setminus \Sigma^{(c)}$, where
``the cut'' $ \Sigma^{(c)}$ is the set $\{(z,z')\in
X^{(c)}\times X^{(c)}: [z\cdot z'] \in [-\infty, -1]\}$. Such a perikernel is
{\sl invariant} if it moreover satisfies in
$\Delta$ the following
 condition:
\begin{equation}
 {\rm W}(gz,gz')= {\rm W}(z,z'), \label{covcom}
\end{equation}
for all $g\in G^{(c)}$.

Since $\Delta =
\{(z,z') \in
X^{(c)}\times X^{(c)}: [z\cdot z'] \in \hat \Theta_L \}$, it follows from
Proposition 3 that ${\rm W}$ is holomorphic in particular in the two tuboids
${\mathcal T}^{- +}= {\mathcal
T}^-\times{\mathcal T}^+$ and
${\mathcal T}^{+ -} = {\mathcal T}^{+}\times {\mathcal T}^{-}$ of
$X^{(c)}\times X^{(c)}$. The corresponding restrictions ${\rm W}^{- +}$ and ${\rm W}^{+ -}$ of ${\rm W}$
admit boundary-values on $X$, denoted respectively
${\mathcal W}^{- +} ={\mathcal W}(x,x')$ and
${\mathcal W}^{+ -}= {\mathcal W}(x',x)$ (the symmetry of these two boundary values being a consequence of
(\ref{covcom})).
These two distributions are such that the difference ${\mathcal C}= {\mathcal W}^{- +} - {\mathcal W}^{+ -}$
(i.e. the discontinuity of ${\rm W}$ across the cut $\Sigma^{(c)}$) vanishes for $[x \cdot x']
> -1 $.

In view of its $ G-$invariance property (\ref{covcom}),
${\rm W}(z,z')$ can also be identified with a function
${\rm w}([z\cdot z'])$ holomorphic in the cut-plane $\hat \Theta_L$.
which can be called a reduced form of ${\rm W}$.

\vskip5pt

Basic examples of invariant perikernels
\footnote{In the application to de Sitter quantum field theory, these examples are interpreted
as the two-point functions (or ``propagators'') of massive Klein-Gordon-like fields \cite{[BM]}
up to an appropriate normalization.}
are provided by the first-kind Legendre functions,
as shown by the following statement
(see also \cite{[BM]} where a
generalization to the $d-$dimensional case is given)
\begin{proposition}
The following integral representation holds:
\begin{equation}
P_{-\frac{1}{2} +i\nu} ([z\cdot z'])=
\frac{e^{\mp\pi \nu}}{\pi}
\int_{\gamma}[z\cdot \xi]^{-\frac{1}{2}
+ i\nu}
[\xi \cdot z']^{-\frac{1}{2} - i\nu}{\rm d}\mu_\gamma(\xi).
\label{legendre}
\end{equation}
In the latter,
the class of integration cycles $\gamma$
and the corresponding measures ${\rm d}\mu_{\gamma}$ are those defined in subsection 4.2
and the integral defines a pair of holomorphic functions
in the respective domains ${\mathcal T}^{- +}$ and ${\mathcal T}^{+ -}$,
corresponding to the choice of sign -- or + in the exponential in front of the integral.
Moreover, formula (\ref{legendre}) defines the first-kind Legendre functions as a family of invariant
perikernels
${\rm W}_{(\nu)}(z,z')=
{\rm w}_{(\nu)}([z\cdot z'])  =
P_{-\frac{1}{2} +i\nu} ([z\cdot z'])$
depending on the parameter $\nu$.
\end{proposition}
\begin{proof}\ \
Call $ {\rm W}^{- +}_{(\nu)}(z,z')$
and $ {\rm W}^{+ -}_{(\nu)}(z,z')$ the pair of functions defined by
the r.h.s. of Eq. (\ref{legendre}) as holomorphic functions (in view of Proposition 1) in the
respective tuboids
${\mathcal T}^{- +}$ and ${\mathcal T}^{+ -}$,

The independence of these functions with respect to the choice of
$\gamma$, which is a consequence of proposition 10 and of Stokes theorem, can be used to show
that they enjoy the $G-$invariance property.
In fact, starting from e.g. $\gamma = \gamma_0$,
one sees that all the cycles $\gamma = g\gamma_0$, with $g \in G$, are homotopic to $\gamma_0$,
and that each measure ${\rm d}\mu_{\gamma}$ is the transform of
${\rm d}\mu_{\gamma_0}$ by the action of $g$ and is itself invariant under the (rotation) subgroup of $G$
which leaves $\gamma$ invariant.
This allows one to make the change of variable $\xi \to g\xi$ for any $g \in G$
in (\ref{legendre}) and therefore
to check that
$ {\rm W}^{\mp \pm}_{(\nu)}(z,z') =
 {\rm W}^{\mp \pm}_{(\nu)}(gz,gz')$ for all $g \in G$.

In order to compute $ {\rm W}^{\mp \pm}_{(\nu)}(z,z')$ explicitly,
we choose $\gamma = \gamma_0$ and a corresponding appropriate
configuration in ${\mathcal T}^{- +}$, namely the pair $(z=
(-i\cosh v, 0, i\sinh v),\ $ $ z' =(i,0,0))$, such that $[z \cdot
z'] = \cosh v \in {\mathbb R}_+$. We then obtain:
\begin{equation}
{\rm w}_{(\nu)}(\cosh v) =
\frac{1}{\pi }
\int_{-\pi }^{\pi} (\cosh v +\sinh v \sin \alpha)^{-\frac{1}{2}
+ i\nu}
\frac{{\rm d}\alpha}{2}
= P_{-\frac{1}{2} +i\nu} (\cosh v).
\label{leg}
\end{equation}
The latter equality results from a standard integral representation of the Legendre function
of first kind \cite{[ER]}. Of course, by taking the complex conjugate configurations, one
would justify similarly the corresponding form of Eq. (\ref{legendre}) valid for
$(z,z') \in {\mathcal T}^{+ -}$.
\end{proof}

\begin{remark}
The function ${\rm W}_{(\nu)}(z,z')$ is a solution in both
variables of the Laplace-Beltrami equation $[\Delta_X - (\nu^2 +
\frac{1}{4})] {\rm W}_{(\nu)}(z,z') =0$ on the manifold $X^{(c)}$
(or in other words, the Klein-Gordon equation on complexified de
Sitter spacetime). For such a $G-$invariant solution, this
equation reduces to a Legendre equation in the variable $[z\cdot
z'] = \cosh v$ and the required analyticity domain $\hat\Theta_L=
{\mathbb C} \setminus ]-\infty,-1]$ selects the first-kind
Legendre function $P_{-\frac{1}{2} +i\nu} (\cosh v)$ as the unique
relevant solution (up to a constant factor).
\end{remark}

We now come back to the Cauchy kernel $[(z-z')^2]^{-1}
= \{-2([z\cdot z']+1)\}^{ -1} $,
whose analyticity in $\hat \Theta_L$ and
$G-$invariance show that it defines an invariant perikernel on $X^{(c)}$:
we prove that the r.h.s. of Eq. (\ref{expcauchy}) is precisely a representation of this
perikernel ${\mathcal K}(z,z')$ in the pair of tuboids
${\mathcal T}^{- +}$ and ${\mathcal T}^{+ -}$.

\vskip 0.4cm
\noindent
{\bf Proof of Proposition 11}

\vskip 0.4cm
\noindent
{\bf A)}  We first give a fast proof which is based
on the integral representation (\ref{legendre}) of the first-kind
Legendre functions
and
on the following Mehler's formula
(\cite{[ER]}, Eq. 3.14.7, taken for $ y=1, p \to -\nu$):
\begin{equation}
\frac{1}{1-x} = \pi \int_0^{\infty} \frac{\nu \tanh \pi \nu}{\cosh \pi \nu}
\ P_{-\frac{1}{2} - i \nu}(-x)\ {\rm d}\nu,
\label{Mehler}
\end{equation}
which yields for $[z\cdot z'] = -x \in \hat\Theta_L$\
\ (since $z^2 = {z'}^2 = -1$):
\begin{equation}
\frac{1}{(z'-z)^2} = - \frac{\pi}{2}
\int_0^{\infty} \frac{\nu \tanh \pi \nu}{\cosh \pi \nu}
\ P_{-\frac{1}{2} - i \nu}([z\cdot z'])\ {\rm d}\nu.
\label{mehler}
\end{equation}
By plugging the expression (\ref{legendre}) of
$P_{-\frac{1}{2} - i \nu}([z\cdot z'])$ into the r.h.s. of (\ref{mehler}), one readily obtains
(\ref{expcauchy}) with the specifications of Proposition 11.

\begin{remark}\ \ It follows from the well-known symmetry property of Legendre functions
$ P_{-\frac{1}{2} - i \nu} =$
$ P_{-\frac{1}{2} + i \nu}, $ that the integration range over $\nu$ can be as well
replaced by $[-\infty, +\infty]$ (together with an extra factor $1/2$) in formula
(\ref{expcauchy}). We shall recover this result and see another aspect of it
below.
\end{remark}

\vskip 0.4cm
\noindent
{\bf B)} We now give an alternative direct proof which exhibits other aspects of
the Cauchy kernel on $X$.

Consider the kernel ${\mathcal K}(z,z')$ defined by the r.h.s. of
Eq. (\ref{expcauchy}) for $z\in {\mathcal T}^-$ and $z' \in {\mathcal T}^+$.
The same argument as in the proof of Proposition 12
shows the $G-$invariance of this kernel and also allows the
use of special configurations, such as
$(z= (-i\cosh v, 0, i\sinh v),$  $ z' =(i,0,0))$ and the integration cycle $\gamma_0$,
for making the computations simpler;
we shall exploit this possibility below.

\vskip 0.3cm
At first, we invert the integrals over $\nu$ and over $\xi$ and thus reexpress ${\mathcal K}$ as follows:
\begin{equation}
{\mathcal K}(z,z') = -\frac{1}{2}\int_{\gamma} {\rm d}\mu_{\gamma}(\xi)
\int^{ +\infty}_{0}[ z\cdot \xi]^{ -{1 \over 2}+i\nu}
[\xi\cdot z']^{ -{1 \over 2}-i\nu}  \nu e^{-\pi \nu}{ {  \tanh} \ \pi \nu
\over {\cosh} \ \pi \nu} \ {\rm d} \nu
\label{defK}
\end{equation}

Let us put: $[ z\cdot \xi] =A(\xi,z) $\nobreak\ \nobreak\ \nobreak\ \nobreak\ $[ z'\cdot
\xi] =B(\xi,z') $\nobreak\ \nobreak\ \nobreak\ \nobreak\ $ a= {\log} \ A,
$\nobreak\ \nobreak\ \nobreak\ \nobreak\ $ b= {\log} \ B $
and
\begin{equation}
w[\xi,z,z'] = (a-b+i\pi )=  {\log} \left({-[z\cdot \xi ] \over [z' \cdot \xi ]}
\right).
\label{w}
\end{equation}
Note that
for $z\in {\mathcal T}^-$ and $z' \in {\mathcal T}^+$, one has
$0 < {\rm Arg} \ B<\pi $
and $ -\pi < {\rm Arg} \ A<0$,
and therefore
(since $ {\rm Im} \ w=\pi + {\rm Arg} \ A- {\rm Arg} \ B$):
$|{\rm Im} \ w| < \pi.$
This implies the convergence and uniform boundedness of the integral
\begin{equation}
J_+(w) =
\int^{ +\infty}_{0} e^{iw\nu} \ \ {\nu\ {  \tanh} \ \pi \nu \over {\cosh} \ \pi
\nu} {\rm d} \nu
\end{equation}
which allows one to rewrite the expression (\ref{defK}) of ${\mathcal K}$ as follows:
\begin{equation}
{\mathcal K}(z,z')= -\frac{1}{2} \int_{\gamma} {\rm d}\mu_{\gamma}(\xi)\ \
{J_+(w[\xi,z,z']) \over ([z\cdot\xi][z'\cdot\xi])^{1/2}}.
\label{newdefK}
\end{equation}

Choosing the previous special configurations for $z,z'$ (with
$[z\cdot z'] = \cosh v$) and $\gamma = \gamma_0$, $w$ becomes a function
$w(\alpha,v)$ of the parameter $\alpha $ of $\gamma_0$, such that
$[z\cdot\xi][z'\cdot\xi]
= \cosh v + \sin \alpha \sinh v = e^w $; this yields the following equivalent forms
for the integral
(\ref{newdefK}):
\begin{equation}
{\mathcal K}(z,z')= -\frac{1}{4}\int^{ \pi}_{-\pi}
{{J_+(w(\alpha,v))}\
{\rm d} \alpha \over( {\cosh} v+ {\sin}
 \alpha \ {  \sinh}  v)^{1/2}}
= -\frac{1}{2}\int^ v_{-v}{
{ J_+(w)}\
{\rm d} w \over[ 2( {\cosh}  v- {\cosh}
w)]^{1/2}}.
\end{equation}
or equivalently:
\begin{equation}
{\mathcal K}(z,z')=
-\frac{1}{4} \int_{-v}^v {J(w)\  {\rm d} w \over[ 2( {\cosh} \ v- {\cosh} \
w)]^{1/2}},
\label{abel}
\end{equation}
where
\begin{equation}
J(w) = J_+(w) + J_+(-w) =
\int^{ +\infty}_{-\infty} e^{iw\nu} {\nu\ {  \tanh} \ \pi \nu \over {\cosh} \ \pi
\nu} {\rm d} \nu
\label{defJ}
\end{equation}
Integration by parts allows one to rewrite Eq. (\ref{defJ}) as follows:
\begin{equation}
J(w)=  \frac{{\rm d}}{{\rm d}w}\left[\frac{w}{\pi}
\int_{-\infty}^{\infty } e^{iw\nu}{ 1 \over {\cosh} \ \pi \nu} {\rm d} \nu \right]
=\frac{1}{\pi}{ {\rm d} \over {\rm d} w} \left[{w \over {\cosh
 \ {w \over 2}}} \right]
\end{equation}
The latter expression has been obtained by applying the residue
method to the intermediate integral (with the cycle $\Gamma$ with
support ${\mathbb R} \cup \{{\mathbb R} + i\}$), namely:
\begin{equation}
\int_{-\infty}^{\infty } e^{iw\nu}{ 1 \over {\cosh} \ \pi \nu} {\rm d} \nu =
\frac{1}{1+e^{-w}}\int^{ }_ \Gamma e^{iw\nu}{ 1 \over {\cosh} \ \pi \nu} {\rm d} \nu =
\frac{ 2e^{-{w/2}}}{1+e^{-w}} = \frac{1}{\cosh {w/2}}.
\end{equation}
We can therefore rewrite the expression (\ref{abel}) of ${\mathcal K}$ as follows:
\begin{equation}
{\mathcal K}(z,z')=-\frac{1}{4\pi} \int^ v_{-v}{ {\rm d} w \over[ 2( {\cosh}  v - {\cosh}
w)]^{1/2}}{ {\rm d} \over {\rm d} w} \left[{w \over {\cosh} \ {w \over 2}}
\right]
\label{Abel}
\end{equation}
We now compute this integral by rewriting it equivalently:
\begin{equation}
{\mathcal K}(z,z')=-\frac{1}{8\pi} \int^ v_{-v}{ {\rm d} w \over
\left[ {\cosh}^2{v \over 2}-
{\cosh}^2{w \over 2} \right]^{1/2}}
\left[{1 \over {\cosh} \ {w \over 2}}-{w
\over 2}{ {  \tanh} \ {w \over 2} \over {\cosh} \ {w \over 2}} \right]
\end{equation}
Taking the variable $ {  \tanh} \ {w \over 2}=t $ and putting $ {  \tanh} \ {v
\over 2}=s $ $ (\vert s\vert \leq 1) $ then yields:
\begin{equation}
{\mathcal K}(z,z')
=-\frac{1}{4\pi} \left(1-s^2
\right)^{1/2} \int^{ +s}_{-s}{ {\rm d}t  \over \left(s^2-t^2 \right)^{1/2}}
\left[1-{t \over 2}\ {\log} \ {1+t \over 1-t} \right]
\end{equation}
or
\begin{equation}
{\mathcal K}(z,z')
=-\frac{1}{4\pi} \left(1-s^2 \right)^{1/2} [\pi - I(s)]
\label{expK}
\end{equation}
with
\begin{equation}
I(s)= \int^ s_{-s}{ {\rm d} t \over \left(s^2-t^2 \right)^{1/2}}\ t\ {\log}
\ (1+t) {\rm d} t.
\end{equation}
Integration by parts yields:
\begin{equation}
I(s)= \int^{ s}_{-s}{  \left(s^2-t^2 \right)^{1/2}
\over 1+t}  {\rm d} t,
\end{equation}
\begin{equation}
{ {\rm d} I(s) \over {\rm d} s}=s \int^{ s}_{-s}{ {\rm d} t \over (1+t)
\left(s^2-t^2 \right)^{1/2}}={s \over 2} \oint^{ }_{ }{ {\rm d} t \over (1+t)
\left(s^2-t^2 \right)^{1/2}}
\end{equation}
In the latter integral, the integration is done on a clockwise contour
surrounding the interval $[-s,s]$ and the residue theorem yields:
\begin{equation}
{ {\rm d} I(s) \over {\rm d} s}
={\pi s \over \left(1-s^2 \right)^{1/2}}.
\end{equation}
Therefore (since $I(0) =0$) we obtain $ I(s)=-\pi \left(1-s^2 \right)^{1/2}+ {\rm \pi} $ and
in view of (\ref{expK}):
\begin{equation}
{\mathcal K}(z,z')=-\frac{1}{4} \left(1-s^2 \right)=-\frac{1}{4 {\cosh}^2{v \over
2}}=\frac{1}{(z-z')^2}
\end{equation}
(A similar computation holds for complex conjugate configurations $(z,z') \in {\mathcal T}^{+ -}$).
\vskip 0.4cm
\noindent
{\sl A geometrical interpretation of the integral (\ref{Abel}):}
\vskip 0.3cm

By taking into account the earlier definition (\ref{w}) of the
variable $w$ as $w[\xi,z,z'] ,$
we see that
the integral over $w$ in (\ref{Abel}) can be transformed back to an integral over $\xi$
similar to (\ref{defK}) or (\ref{newdefK})
on a general cycle $\gamma$, but in which the integral over $\nu$
is equal to $J(w)$ instead of $J_+(w)$
(corresponding to the integration range $\nu \in [-\infty, +\infty]$).
This yields:
\begin{equation}
{\mathcal K}(z,z') = -\frac{1}{4\pi}
\int_{\gamma} \frac{{\rm d}\mu_{\gamma}(\xi)}{\left([z\cdot \xi][ z'\cdot \xi] \right)^{1/2}}
{{\rm d} \over {\rm d} w} \left[{w \over {\cosh} \ {w \over 2}} \right](\xi,z,z')
\label{precaufan}
\end{equation}
and justifies
the following alternative form to Proposition 11
\begin{proposition}
The Cauchy kernel on $X^{(c)}$ admits the following alternative expressions
\begin{equation}
\frac{1}{(z'-z)^2} = -  \frac{1}{4}
\int_{-\infty}^{\infty}
\frac{\nu \tanh \pi \nu}{e^{\pm \pi\nu} \cosh \pi \nu} {\rm d}\nu
\int^{ }_ \gamma
[z\cdot \xi]^{-{1 \over 2}+i\nu}
[\xi\cdot z']^{-{1 \over 2}-i\nu}
{\rm d}\mu_{\gamma} (\xi)
\end{equation}
\begin{equation}
{\  }= -\frac{1}{4\pi i}
\int_{\gamma} {\rm d}\mu_{\gamma}(\xi)
\left[ \frac{1}{[(z-z')\cdot \xi]} - \frac{
\log \left(\frac{ - [z \cdot\xi]}{[z' \cdot \xi]}\right) \ [(z+z')\cdot \xi]}
{[(z-z')\cdot\xi]^{2} } \right],
\label{caufan}
\end{equation}
valid for all $(z,z')$ in ${\mathcal T}^{- +}$ (resp. in
${\mathcal T}^{+ -}$).
\end{proposition}
We notice that the latter expression (\ref{caufan}) of the Cauchy kernel
(obtained from Eq. (\ref{precaufan}) by replacing $w$ by its expression (\ref{w})) has the form
of a Cauchy- Fantappi\'e-like kernel, interpretable as an integral over a set of
analytic singularities (namely the hyperplanes with equation $[(z-z')\cdot \xi] = 0$)
which are exterior, but tangential, to the holomorphy domains
${\mathcal T}^{- +}$ and
${\mathcal T}^{+ -}$ of $X^{(c)}$.
\begin{remark}
The inversion formula (\ref{invfourier}) given in Theorem 2 can equivalently be replaced by the
following one
\begin{equation}
F(z) =  \frac{1}{4\pi^2} \int_{-\infty}^{\infty}
\frac{\nu \tanh \pi \nu}{e^{\pm\pi\nu} \cosh \pi \nu} {\rm d}\nu
\int^{ }_ \gamma
[z\cdot \xi]^{-{1 \over 2}+i\nu} \tilde{f}_{\pm,\nu}
(\xi)
{\rm d}\mu_ \gamma (\xi)
\label{invfourier2}
\end{equation}
although the equality of the subintegrals over $\nu$ on $[0,+\infty[ $ and $]-\infty,0]$
is not a trivial symmetry of the formula, but results from the symmetry
$\nu \to -\nu$ which appeared in the computation of the Cauchy kernel (see Remark 4).
In fact, this shows that the knowledge of $F(z)$ (for e.g. $z \in {\mathcal T}^{- }$)
is entirely encoded in the knowledge of
$\tilde f_{+,\nu}$ for all positive $\nu$, or for all negative $\nu$, although
$\tilde f_{+,\nu}$ and
$\tilde f_{+,-\nu}$ are not related in a simple way.
\end{remark}

\section{Fourier-Helgason transform of $G_b-$invariant functions and spherical Laplace transform
of Volterra kernels}
Our aim is to apply the previous definition (\ref{Fourier}) of the Fourier-Helgason transformation
to a  particular class of
boundary values of holomorphic functions in the tuboids ${\mathcal T}^+$ and ${\mathcal T}^-$:
these functions, denoted $\underline{\rm W}^{\pm}(z)$, enjoy the additional property of being invariant under
the stabilizer subgroup $G^{(c)}_b$ of the base point $b =(0,0,1)$.

Let us recall
(see \cite{[BV]} and
\cite{[B]})
how such functions occur as representatives of
invariant perikernels
and at first what is the relationship of the latter with the invariant Volterra kernels on $X$
introduced in \cite{[FV]}, \cite{[Fa]}.

\vskip5pt
\noindent
1)\  Let ${\rm W}(z,z') = w([z\cdot z'])$ be any invariant perikernel, holomorphic in the domain
$\Delta$ of $X^{(c)} \times X^{(c)}$ (see subsection 4.3),
${\mathcal W}^{- +}$
and ${\mathcal W}^{+ -}$ its boundary values on $X \times X$ from the respective tuboids
${\mathcal T}^{- +}$ and
${\mathcal T}^{+ -}$. Their difference
${\mathcal C}= {\mathcal W}^{- +}- {\mathcal W}^{+ -}$ has its support contained in the
set $\{(x,x')\in X\times X;\ (x-x')^2 \geq 0\}. $
This set admits a splitting into two $G-$invariant closed parts,
having only in common the diagonal set
$\{(x,x'); x =x' \}$ and
characterized by
the respective conditions $x \geq x'$ and
$x' \geq x$
($\geq$ being the order relation on $X$ defined at
the beginning of section 2).
One can thus write a decomposition of ${\mathcal C}$ of the form
${\mathcal C} = -i\left( {\mathcal R} - {\mathcal A}\right)$, such that the support of ${\mathcal R}$
(resp. of ${\mathcal A}$) is
contained in the set
$\{(x,x')\in X\times X;\  x \geq x' \}$
(resp. $\{(x,x')\in X\times X;\  x' \geq x \}$).
{\sl ${\mathcal R}$ (and also ${\mathcal A}$) is then a
Volterra kernel on $X$ in the sense of \cite{[Fa]}, which is associated with the perikernel
${\rm W}$.} In view of the time-ordering interpretation of the relation $x \geq x'$,
${\mathcal R}$ and ${\mathcal A}$ are called respectively ``retarded '' and ``advanced'' kernels.

\vskip5pt
\noindent
2)\ By using the transitivity of the group
$G^{(c)}$ on  $X^{(c)}$,  one can then identify
${\rm W}(z,z')$ with a $G^{(c)}_b$-invariant function
$\underline{\rm{W}}(z)= {\rm W}(z,b)$;
$\underline{\rm{W}}$ is
holomorphic in the domain
\begin{equation}
\underline{\Delta} = \left\{ z\in X^{(c)}, \; z_2 = -[z\cdot b]
\in {{\mathbb C}}\setminus [1,\infty[\right\}.
\end{equation}
In view of its $ G^{(c)}_b-$invariance property,
$\underline{\rm{W}}(z)$ can also be identified with the function
${\rm w}(-z_2)$ holomorphic in the cut-plane $\hat \Theta_L =
{\mathbb C}\setminus ]-\infty, -1]$. $\underline{\rm W}$, like
${\rm w}$, is a  reduced form of ${\rm W}$. It follows from
Proposition 3 (since $\hat \Theta_L \supset \Theta_L$) that
$\underline{\rm W}$ admits restrictions to the tuboids ${\mathcal
T}^{-} $ and  ${\mathcal T}^{+}$, which we call respectively
$\underline{\rm W}^-$ and $\underline{\rm W}^+$. The corresponding
boundary values on $X$, denoted $\underline{\mathcal W}^{\pm},$
are such that $\underline{\mathcal W}^-(x)$ =${\mathcal W}^{-
+}(x,b)$ and $\underline{\mathcal W}^+(x)$ =${\mathcal W}^{+
-}(x,b)$. The corresponding jump of $\underline{\rm W}$, namely
the function $\underline{\mathcal C} = \underline{\mathcal W}^-
-\underline{\mathcal W}^+$ (such that  $\underline{\mathcal C}(x)
= {\mathcal C}(x,b)$), has its support contained in
$\Gamma^{+}(b)\cup\Gamma^{-}(b) = \{ x \in X;\;x_2 \geq 1\}$. The
boundary values $\underline{\mathcal W}^+$ and
$\underline{\mathcal W}^-$ therefore coincide in the open set
${\mathcal U}_{b}= \{x\in X;\;x_2 < 1\}$, being in fact identical
with the restriction of the analytic function  $\underline{\rm W}$
to ${\mathcal U}_{b}$. Similarly, {\sl the associated  invariant
Volterra kernel ${\mathcal R}$ (resp. ${\mathcal A}$) admits a
reduced form} $\underline{\mathcal R}(x)={\mathcal R}(x,b) =
r(x_2)Y(x_0)$ (resp. $\underline{\mathcal A}(x)={\mathcal A}(x,b)
= a(x_2)Y(-x_0)$). The latter correspond to a decomposition
$\underline{\mathcal C} = -i\left( \underline{\mathcal R} -
\underline{\mathcal A}\right)$ of $\underline{\mathcal C}$, such
that the supports of $\underline{\mathcal R}$ and
$\underline{\mathcal A}$ are respectively contained in the sets
$\Gamma^{+}(b)$ (i.e. $x_2 \geq 1,\ x_0 \geq 0$) and
$\Gamma^{-}(b)$ (i.e. $x_2 \geq 1,\ x_0 \leq 0$) (these are the
``future cone'' and ``past cone'' of $b$ in $X$). The
corresponding function $r(x_2)$ has therefore its support
contained in $[1,+\infty[$; $\underline{\mathcal R}$ and $ r$
(resp. $\underline{\mathcal A}$ and $a$) are also called the
retarded (resp. advanced) functions associated with
$\underline{\rm W}$; moreover, {\sl the $G_b-$invariance of
$\underline W$ implies the equality $a(x_2) = r(x_2)$.}

\subsection{The spherical Laplace transformation as a
special case of the Fourier-Helgason transformation.}
We shall make use of the following
\begin{proposition}
If a function $f$ on $X$ is invariant under the subgroup $G_b$ of $G$, then
its Lorentzian FH-transforms are of the following form:
\begin{equation}
\tilde f_{\pm,\nu}(\xi) = \left|[\xi\cdot b]\right|^{-\frac{1}{2} -i\nu}
\left(Y([\xi\cdot b])
\tilde F_{\pm}(\nu)
+ Y(-[\xi\cdot b] ) \tilde F'_{\pm}(\nu)\right)\
\label{FHGinv}
\end{equation}
\end{proposition}
\begin{proof} It is clear from the defining formula (\ref{Fourier}) that if $f(x)$ is invariant under
$G_b$, then $\tilde f_{\pm,\nu}(\xi)$ is also invariant under $G_b$, considered as acting on the cone
$C^+$. Now there are two classes of orbits of $G_b$ on $C^+$, which are generated respectively by the
action of the positive dilatations $\xi \to r \xi, \ r >0,$ on the following special orbits:
$\omega = \{\xi = (\cosh \chi, \sinh \chi,-1)\}$ and
$\omega' = \{\xi = (\cosh \chi, \sinh \chi,1)\}$.
Then (provided they are defined)
the Lorentzian FH-transforms of $f$ are constant
on these orbits $\omega$ and $\omega'$,
namely one can put $ \left.\tilde f_{\pm,\nu}\right|_{\omega} = \tilde F_{\pm}(\nu)$ and
$ \left. \tilde f_{\pm,\nu}\right|_{\omega'} = \tilde F'_{\pm}(\nu).$
Then for all $\xi \in C^+$, with $[\xi \cdot b] \neq 0$, the form (\ref{FHGinv}) of
$\tilde f_{\pm,\nu}(\xi)$ follows from its homogeneity property, and this is
sufficient for
defining
$\tilde f_{\pm,\nu}(\xi)$ as an $L^1-$function of $\xi$ on each cycle $\gamma$ equipped
with the measure ${\rm d}\mu_{\gamma}$.
We moreover see that the functions $\tilde F_{\pm}$ and $\tilde F'_{\pm}$ can be computed simply by fixing
two special configurations of $\xi$, namely one has:
$$ \tilde F_{\pm}(\nu)  =  \tilde f_{\pm,\nu}((1,0,-1))$$
and
$$ \tilde F'_{\pm}(\nu)  =  \tilde f_{\pm,\nu}((1,0,1)).$$
\end{proof}

We shall  now consider a system of $G_b-$invariant functions
$\underline{\mathcal W}^-,$
$\underline{\mathcal W}^+,$
$\underline{\mathcal C},$
$\underline{\mathcal R},$
$\underline{\mathcal A},$
associated with an invariant perikernel ${\rm W}$ and compute the Lorentzian FH-transforms
of these various functions.
We shall see that it is sufficient to make the computation for the retarded function
$\underline{\mathcal R}(x)=r(\cosh v)$: here,we have taken into account the support of $r$ by
putting $x_2 = \cosh v$, with $v \geq 0$.
In view of Proposition 14, we are led to compute the corresponding quantities
$ \tilde F_{\pm}(\nu)  \equiv  \tilde {\underline{\mathcal R}}_{\pm,\nu}((1,0,-1))$
and
$ \tilde F'_{\pm}(\nu)  \equiv  \tilde {\underline{\mathcal R}}_{\pm,\nu}((1,0,1)).$
\vskip 0.3cm

To this purpose we introduce the following transforms $G$ and $H$ of $r$:
\begin{equation}
G(\nu) = \int_{0}^{\infty}
Q_{-\frac{1}{2} +i\nu}
(\cosh v)\  r(\cosh v) \ \sinh v\  {\rm d}v,
\label{qtrans}
\end{equation}
\begin{equation}
H(\nu) = \int_{0}^{\infty}
P_{-\frac{1}{2} +i\nu}
(\cosh v)\  r(\cosh v)\
\sinh v\  {\rm d}v.
\label{ptrans}
\end{equation}
In the latter
$P_{-\frac{1}{2} +i\nu}$
and $Q_{-\frac{1}{2} +i\nu}$
denote respectively the first-kind and second -kind Legendre functions,
and the well-known identity \cite{[ER]} $Q_{-\frac{1}{2} +i \nu} -
Q_{-\frac{1}{2} -i \nu} =
-i \pi \tanh \pi \nu\  P_{-\frac{1}{2} +i \nu} $ implies
the following relation:
\begin{equation}
H(\nu) = H(-\nu) = \frac {G(\nu)-G(-\nu)} {- i \pi \tanh \pi \nu}
\label{rel}
\end{equation}

We can then show the following property:
\begin{proposition}
Under the assumption that $e^{\frac{v}{2}} r(\cosh v)$ is in $L^1
({\mathbb R}^+,{\rm d}v)$, the Lorentzian FH-transforms of
$\underline{\mathcal R}$ are obtained by the following formulae:
\begin{equation}
\tilde {\underline{\mathcal R}}_{\pm,\nu}(\xi) = \left|[\xi\cdot b]\right|^{-\frac{1}{2} -i\nu}
\left(Y([\xi\cdot b])
\tilde F(\nu)
+ Y(-[\xi\cdot b]) \tilde F'_{\pm}(\nu)\right)\
\label{FHRinv}
\end{equation}
where
\begin{equation}
a)\ \ \ \ \ \tilde F(\nu)  \equiv  \tilde {\underline{\mathcal R}}_{\pm,\nu}((1,0,-1))
=  G(\nu);
\label{FRG}
\end{equation}
$G(\nu)$ coincides
\footnote{up to the change of variable $\nu \to \lambda =-\frac{1}{2} + i\nu$
and a normalization factor equal to $1/2$: see e.g. Eqs (III.10),(III.11) of \cite{[BV]}.}
with the spherical Laplace transform \cite{[FV]} \cite{[BV]} of the Volterra kernel ${\mathcal R}(x,x')$ whose
reduced form is $\underline {\mathcal R}$; $G(\nu)$ is  the boundary value
of a holomorphic function in the
half-plane ${\rm Im} \nu <0$.
\begin{equation}
b)\ \   \tilde F'_{\pm}(\nu)  \equiv  \tilde {\underline{\mathcal R}}_{\pm,\nu}((1,0,1))
=  \frac{\pi H(\nu)}{\cosh \pi \nu}  \mp i e^{\pm \pi \nu} G(-\nu)
\label{FRH}
\end{equation}
\end{proposition}
\begin{proof}
a) \  Let $\xi = (1,0,-1)$; since the support $\Gamma^+(b)$ of
$\underline{\mathcal R}$ is contained in the region where $[x\cdot \xi] = x_0 + x_2 $ is positive,
we can put $[x\cdot \xi] = e^t$ and use the following parametrization of $\Gamma^+(b)$ in
terms of $t$ and $v$:
\begin{equation}
\Gamma^+(b) \ \ \ \ \left\{
\begin{array}{ccl}
x_0 & = &  e^t - \cosh v\\
x_1 & = & \pm  e^{t/2}\sqrt{2(\cosh t - \cosh v)}\\
x_2 & = &   \cosh v\\
\end{array}
\right.
\ \ \ \makebox{with}\;\; t \geq v \geq 0.
\end{equation}
In these coordinates one has
\begin{equation}
{\rm d}\sigma(x) = \frac{e^{t/2}\sinh v\  {\rm d}t\ {\rm d}v}{2 \sqrt{2(\cosh t - \cosh v)}}
\label{sigma}
\end{equation}
We therefore have in view of (\ref{Fourier}) (since
$(x_0 + x_2)_{\pm}^{-\frac{1}{2} - i\nu} \equiv
(x_0 + x_2)^{-\frac{1}{2} - i\nu} $ in this region):
\[
\tilde F(\nu) =
\tilde {\underline{\mathcal R}}_{\pm,\nu}((1,0,-1)) =
\int_{\Gamma^+(b)}(x_0 + x_2)^{-\frac{1}{2} - i\nu} r(x_2)\  {\rm d}\sigma(x)
   \]
\begin{equation}
=2 \int_{0}^{\infty}r(\cosh v)\  \sinh v\  {\rm d}v
 \int_{v}^{\infty}
 \frac
 {e^{-i \nu t }
 \ {\rm d}t}
{2 \sqrt{2(\cosh t - \cosh v)}}
 =  G(\nu);
\label{FHRG}
\end{equation}
For obtaining the latter, we have
taken into account our assumption on $r$ which ensures the convergence of the double integral and
we have made use of the following
integral representation of the second-kind Legendre
function
\begin{equation}
Q_{-\frac{1}{2} +
i\nu}(\cosh v)
 =\int_{v}^{\infty}
 \frac
 {e^{-i \nu t }
 \ {\rm d}t}
{\sqrt{2(\cosh t - \cosh v)}}
\label{seclegendre}
\end{equation}
We notice that in this configuration $\xi = (1,0,-1)$, the two FH-transforms
$\tilde {\underline{\mathcal R}}_{+, \nu}$
and $\tilde {\underline{\mathcal R}}_{-, \nu}$ of $\underline {\mathcal R}$
coincide (as functions of $\nu$) with
the spherical Laplace transform of the Volterra kernel ${\mathcal R}$
(see our previous footnote and
\cite{[FV]}, \cite{[BV]} for a detailed study of this transform).
Under our assumption on $r$, the integral in Eq. (\ref{FHRG}) is uniformly convergent
for all complex $\nu$ such that ${\rm Im}\nu <0$ and therefore extends the definition
of $G(\nu) $ as a
holomorphic function in this domain.
\vskip 0.4cm

b) \  Let $\xi = (1,0,1)$; the support $\Gamma^+(b)$ of
$\underline{\mathcal R}$ is now decomposed into two regions according to the sign of
the scalar product $[x\cdot \xi] = x_0 - x_2. $  The
corresponding FH-transforms
$\tilde {\underline{\mathcal R}}_{\pm,\nu}((1,0,1)) $
are then given as the sum of two contributions

i) In the region
(I)$  =  \Gamma^+(b)\cap\{ x: \,x_0 - x_2 >0\}$:
we can put $[x\cdot \xi] = e^t$ and use the following parametrization of this region
\begin{equation}
({\rm I})\ \ \ \
\left\{
\begin{array}{ccl}
x_0 & = &  e^t + \cosh v\\
x_1 & = & \pm   e^{t/2}\sqrt{2(\cosh v + \cosh t)}\\
x_2 & = &   \cosh v\\
\end{array}
\right.\ \ \  \makebox{with}\;\; t \in {\mathbb R}, \;\;v\geq 0.
\end{equation}
In these coordinates one has
\begin{equation}
{\rm d}\sigma(x) = \frac{e^{t/2}\sinh v\  {\rm d}t\ {\rm d}v}{2 \sqrt{2(\cosh v + \cosh t)}}
\end{equation}
and therefore:
\[
\int_{(\rm I)}(x_0-x_2)^{-\frac{1}{2} -i \nu}_\pm r(x_2) {\rm d}\sigma(x) =
2 \int_{0}^{\infty}
r(\cosh v) \sinh v\,{\rm d}v
\int^{\infty}_{-\infty}
\frac
{ e^{-i \nu t } \ {\rm d}t}
{2 \sqrt{2(\cosh v + \cosh t)}}
\]
\begin{equation}
= \frac{\pi}{ \cosh \pi \nu}
\int_{0}^{\infty}  P_{-\frac{1}{2}  +
 i \nu} (\cosh v)\
 r(\cosh v)\  \sinh v\,{\rm d} v
= \frac{\pi}{ \cosh \pi \nu}  H(\nu).
\label{FHRI}
\end{equation}
\vskip 0.4cm

ii) In the region (II) $ = \Gamma^+(b)\cap \{x: x_0 - x_2 <0 \}$
we can put $[x\cdot \xi] = -e^{t}, $  being careful that
$(x_0-x_2)_{\pm}^{-1/2 -i\nu} =
\mp i e^{\pm \pi \nu}
e^{(-\frac{1}{2}-i\nu) t},$
and use the following parametrization of this region
\begin{equation}
({\rm II})\ \ \ \
\left\{
\begin{array}{ccl}
x_0 & = &  -e^t + \cosh v\\
x_1 & = &  \pm  e^{t/2}\sqrt{2(\cosh t - \cosh v)}\\
x_2 & = &   \cosh v\\
\end{array}
\right. \makebox{for}\;\; t \leq -v \;\;v\geq 0.
\end{equation}
Since ${\rm d}\sigma(x)$ is again given by Eq. (\ref{sigma}),
the contribution of the region II is:
\[\int_{(\rm II)}(x_0 - x_2)^{-\frac{1}{2} - i \nu}_\pm\  r(x_2)
{\rm d}\sigma(x) = \]
\[ \mp 2 i e^{\pm \pi\nu}\int_{0}^{\infty}r(\cosh v) \sinh v \ {\rm d}v
 \int_{-\infty}^{-v}
  \frac{e^{-i \nu t } {\rm d}t}{2 \sqrt{2(\cosh t - \cosh v)}}
\]
\begin{equation}
=\mp  i e^{\pm  \pi\nu)}\int_{0}^{\infty}
Q_{-\frac{1}{2} -
i\nu}(\cosh v)\  r(\cosh v)\  \sinh v\ {\rm  d}v
= \mp i e^{\pm \pi \nu} \ G(-\nu).
\label{FHRII}
\end{equation}
Regrouping together the two contributions (\ref{FHRI}) and (\ref{FHRII}), we then
obtain:
\begin{equation}
\tilde {\underline{\mathcal R}}_{\pm,\nu}((1,0,1))
= \frac{\pi H(\nu)}{\cosh \pi \nu}  \mp i e^{\pm \pi \nu} G(-\nu) ,
\label{FHRtot}
\end{equation}
which ends the proof of b).
\end{proof}

\vskip 0.4cm
Let us rewrite as follows the result of Proposition 15:
\begin{equation}
\tilde{\underline{\mathcal R}}_{\pm, \nu}(\xi) =
\left|[\xi\cdot b]\right|^{-\frac{1}{2} - i\nu}
\left\{
Y ([\xi\cdot b]) G(\nu)
+ Y (-[\xi\cdot b])
\left[\frac{\pi H(\nu)}{\cosh\pi\nu}
\mp i e^{\pm \pi\nu}
G(-\nu)\right]
\right\}
 \label{FHR}
\end{equation}
By the same analysis, a similar formula can be obtained for the
FH-transforms of the corresponding advanced function $\underline{\mathcal A}(x):$
$$ \tilde{\underline{\mathcal A}}_{\pm, \nu}(\xi) =
\left|[\xi\cdot b]\right|^{-\frac{1}{2} - i\nu} \times \cdots  $$
\begin{equation}
\cdots  \left\{
\mp iY (-[\xi\cdot b])  e^{\pm \pi \nu} G(\nu)
+ Y ([\xi\cdot b])
\left[\mp i e^{\pm \pi\nu} \frac{\pi H(\nu)}{\cosh\pi\nu}
+G(-\nu)\right]\right\}
 \label{FHA}
\end{equation}
Taking into account Eqs (\ref{FHR}) and (\ref{FHA}), we now deduce the
FH-transforms of the function
$\underline{\mathcal C} = -i
\left(\underline{\mathcal R} -
\underline{\mathcal A}\right);$ in view of (\ref{rel}), we obtain:
\begin{equation}
\tilde{\underline{\mathcal R}}_{\pm, \nu}(\xi) -
\tilde{\underline{\mathcal A}}_{\pm, \nu}(\xi) =
\pm i
\left|[\xi\cdot b]\right|^{-\frac{1}{2} - i\nu}
{\pi H(\nu)}
\left\{
 e^{\pm i\pi\left(-\frac{1}{2} - i\nu\right)  }
Y (-[\xi\cdot b])
+Y ([\xi\cdot b])
\right\}
\label{FHRA}
\end{equation}
or
\begin{equation}
\tilde{\underline{\mathcal C}}_{\pm,\nu} =
\pm \pi H(\nu)
[\xi\cdot b]^{-\frac{1}{2} - i\nu  }_{\pm}.  \label{FHC}
\end{equation}
Since Proposition 8 can be applied to the holomorphic functions $\underline{\mathcal W}^{\pm}$,
the FH-transforms of
$\underline{\mathcal C} =
\underline{\mathcal W}^-  -
\underline{\mathcal W}^+$ immediately yield those of
$\underline{\mathcal W}^-$ and
$\underline{\mathcal W}^+$, namely:
\begin{equation}
\tilde{\underline{\mathcal W}}^-_{+,\nu}(\xi) =
{ \pi H(\nu)}
[\xi\cdot b]_+^{-\frac{1}{2} - i\nu  },\;\;\;\;
\tilde{\underline{\mathcal W}}^-_{-,\nu}(\xi) =0,
\label{w-tilde}
\end{equation}
\begin{equation}
\tilde{\underline{\mathcal W}}^+_{-,\nu}(\xi) =
{ \pi H(\nu)}
[\xi\cdot b]_-^{-\frac{1}{2} - i\nu  },\;\;\;\;
\tilde{\underline{\mathcal W}}^+_{+,\nu}(\xi) =0,
\label{w+tilde}
\end{equation}

\subsection{Connection between the inverse transformations}
Let us restrict our attention to the function $\underline{\rm W}^-(z)$ and write the inversion formula
(\ref{invfourier}) which expresses it in terms of its FH-transform
$\tilde{\underline{\mathcal W }}^-_{+,\nu}(\xi)
= \pi H(\nu) [\xi\cdot b]_+^{-{1 \over 2}-i \nu}. $
We obtain:
\[
\underline{\rm W}^-(z) =
{\rm w}([z\cdot b]) =
\frac{1}{2\pi^2} \int_0^{\infty}
\frac{\nu \tanh \pi \nu}{e^{\pi\nu} \cosh \pi \nu}\ {\rm d}\nu
\int^{ }_ \gamma
[z\cdot \xi]^{-{1 \over 2}+i\nu}\ \
\tilde{\underline{\mathcal W}}^-_{+,\nu}(\xi)
\ \ {\rm d}\mu_\gamma (\xi) \]
\begin{equation}
 = \frac{1}{2 \pi} \int_0^{\infty}
\frac{\nu \tanh \pi \nu}{e^{\pi\nu} \cosh \pi \nu} \ \ H(\nu)
\ \ {\rm d}\nu
\int^{ }_ \gamma
[z\cdot \xi]^{-{1 \over 2}+i\nu}
[\xi\cdot b]_+^{-{1 \over 2}-i\nu}\ {\rm d}\mu_ \gamma (\xi)
\label{klmno}
\end{equation}
But in the latter we recognize the integral representation (\ref{legendre}) of the
first-kind Legendre function, which therefore allows us to write:
\begin{equation}
{\rm w}([z\cdot b]) =
 \frac {1}{2}  \int_0^{\infty}
\frac{\nu \tanh \pi \nu}{\cosh \pi \nu}
\ \  P_{-\frac{1}{2} - i \nu}([z\cdot b])
\ \ H(\nu)
\ \ {\rm d}\nu
\label{klmn}
\end{equation}
Taking into account Eq.(\ref{rel}), we can at first transform the previous integral over $\nu$
into an integral on the full real axis (with a factor $1/2$) and then purely replace
$\tanh \pi \nu\  H(\nu)$ by $2 \left(\frac{i}{\pi}\right) G(\nu)$ in the integrand
since the remaining factor is an odd function of $\nu$. We thus obtain
the following statement which is a special case of a general property
established in \cite{[BV]}, according to which
a general perikernel with moderate growth (namely dominated by $|[z\cdot z']|^m$, with $m > -1$)
can be decomposed linearly on the family of elementary perikernels $P_{m+i \nu}([z\cdot z'])$,
with $\nu$ varying from $-\infty $ to $+\infty$.

\vskip 0.5cm
\noindent
{\bf Theorem 3}
{\sl Let $W(z,z') ={\rm w}([z\cdot z'])$ be an invariant perikernel on $X$,
whose associated retarded Volterra kernel $R$
satisfies the
assumption of Proposition 15. Then it
admits the following decomposition in terms of
the spherical Laplace transform $G(\nu)$ of $R$:
\footnote{In the context of de Sitter quantum field theory, this result has been interpreted in
\cite{[BM]} as a
K\"allen-Lehmann-type
representation for the two-point functions of
general interacting fields.}
\begin{equation}
{\rm w}([z\cdot z']) =
 \frac{i}{2\pi} \int_{-\infty}^{\infty}
\frac{\nu
\  P_{-\frac{1}{2} - i \nu}([z\cdot z'])}
{\cosh \pi \nu}
\ \ G(\nu)
\ \ {\rm d}\nu.
\label{klm}
\end{equation}}

\section{The chiral Fourier-Helgason transformation}
\subsection{Definition and properties of the transforms
$\tilde f_{\rightarrow}$
and $\tilde f_{\leftarrow}$}
Let us first show the following
\begin{proposition}
For every function $ f $ in $H^2_{(reg)}({\mathcal T}_{\rightarrow})$
or in $H^2_{(reg)}({\mathcal T}_{\leftarrow}) $),
the corresponding Lorentzian FH-transforms
$\tilde{{f}}_{+, \nu}(\xi) $ and  $\tilde{{f}}_{-,\nu}(\xi )$ both vanish.
\end{proposition}
\begin{proof}
It is sufficient to consider the case when $f$ belongs to
$H^2_{(reg)}({\mathcal T}_{\leftarrow})$: the corresponding
function  $\hat f(\lambda, \mu)$ is the boundary value of a
function $\hat F(\lambda, \mu)$ holomorphic in the pierced tube
$\tau^{+,+} \setminus \delta$ (see Proposition 7), which moreover
satisfies the boundedness property of Definition 1. We then
consider the following integral, similar to the one at the r.h.s.
of Eq.(\ref{fourier2}), but taken on a cycle of the form ${\mathbb
R}^2 + i (a \lambda_0, a \mu_0)$ with $a \geq 0, \ 0 < \lambda_0 <
\mu_0$:
\begin{equation}
2^s  \int_{{\mathbb R}^2 + i (a \lambda_0, a \mu_0)}
\left[\frac{\hat F (\lambda,\mu)}{\lambda - \mu}\right]
\frac{[(\cos \alpha/2) \lambda - (\sin \alpha/2)]^s [(\cos
\alpha/2) \mu-(\sin \alpha/2)]^s} {(\lambda - \mu)^{s+1}} {\rm
d}\lambda {\rm d}\mu. \label{fourier4}
\end{equation}
In the latter, each complex power of the form $t^s$
is fixed in its principal sheet (see subsection 4.1).
In view of the analyticity and boundedness properties of the integrand, the latter integral is
well-defined for $-1 < {\rm Re}s < 0, $ and it is independent
of $a$ (since the cycle remains in the tube $\tau^{+,+} \cap \{(\lambda,\mu); \ {\rm Im} (\lambda -\mu)
<0\} $); moreover, it admits a bound which tends to zero when $a$ tends to infinity. Therefore it
vanishes for all $a >0$. Now one checks that since the algebraic factor in the integrand is equal to
$ \left(\cos^2 \frac{\alpha}{2}\right)^s  (\lambda_{\alpha} - \mu_{\alpha})^{-s} , $
the limit of the integral (\ref{fourier4})
for $a$ tending to zero coincides with the r.h.s. of Eq. (\ref{fourier3}) {\sl with the specification
$(\lambda_{\alpha} - \mu_{\alpha})_-^{s+1}$}. This shows that
$ \tilde f_+(\xi(\alpha), s) =0$ (for all values of $\alpha$).
Choosing the previous cycle of integration in
the tube $\tau^{+,+} \cap \{(\lambda,\mu); \ {\rm Im} (\lambda -\mu) >0 \} $ (i.e. with
$ 0 < \mu_0 <  \lambda_0 $), one shows similarly that
$ \tilde f_-(\xi, s) =0.$
\end{proof}

\vskip 0.2cm In order to introduce the chiral FH-transformation,
we need the following geometrical property of the complex cone
$C^{(c)} = \{\xi = (\xi_0,\xi_1,\xi_2)\in {\mathbb C}^3;\ \xi_0^2
-\xi_1^2 -\xi_2^2 =0\}$:
\begin{proposition}
All complex points $\xi \in C^{(c)}$ are such that $({\rm Im}\xi)^2 \leq 0$ and
$C^{(c)}$ admits the following partition:
$C^{(c)} = C \cup C_{\rightarrow} \cup C_{\leftarrow},$ where
$$C_{\rightarrow} = \{ \xi \in C^{(c)};\ \epsilon(\xi) = -\}, $$
$$C_{\leftarrow} = \{ \xi \in C^{(c)};\ \epsilon(\xi) = +\}, $$
with $\epsilon(\xi) =\  {\rm sgn}\  {\rm Det}(e,{\rm Re}\xi,{\rm Im}\xi)$ (see subsection 2.2).

The domains
$C_{\rightarrow}$ and $ C_{\leftarrow},$ (similar to the tuboids
${\mathcal T}_{\rightarrow}$
and ${\mathcal T}_{\leftarrow} $ of $X^{(c)}$) are determined by their bases
in the complex circle $\gamma_0^{(c)}$ of $C^{(c)}$, parametrized by two half-planes:
$$(\gamma_0^{(c)})_{\rightarrow} = \{\xi =\xi(\Phi)= (1, \sin \Phi, \cos \Phi);\
\Phi = \phi + i \eta,\  \eta >0 \}$$
$$(\gamma_0^{(c)})_{\leftarrow} = \{\xi =\xi(\Phi)= (1, \sin \Phi, \cos \Phi);\
\Phi = \phi + i \eta,\  \eta <0 \}$$
\end{proposition}

In the same spirit as for the Lorentzian FH-transformations (see proposition 8 and Definition 2),
one could introduce the chiral transformations by specifying boundary value prescriptions
$[x_{\rightarrow}\cdot \xi]^s$
and $[x_{\leftarrow}\cdot \xi]^s$ for the FH-kernel, corresponding to limits
of the holomorphic function $[z\cdot \xi]^s$ from the respective tuboids
${\mathcal T}_{\rightarrow}$
and ${\mathcal T}_{\leftarrow} $ of $X^{(c)}$
and put correspondingly:

$$\tilde{f}_{\leftrightarrows}(\xi,s) = \int_{X}
[x_{\leftrightarrows}\cdot \xi]^{s} f(x)\ {\rm d}\sigma(x). $$

However, two new features appear in this case:

1) The non-uniformity of the function
$[z\cdot \xi]^s$  for general $s$ leads one to define the transform {\sl only for $s$ integer and negative}
(for the sake of convergence).
(Note that in the $(\lambda,\mu)-$ representation of $X^{(c)}$ (see section 3), this non-uniformity
is easily seen to be due (see Eq. (\ref{zxi}))
to the non-trivial homotopy of the pierced tubes $\tau^{\pm,\pm} \setminus \delta$
which represent
${\mathcal T}_{\leftrightarrows}$).

2) Instead of considering $x$ as the limit of points in $X^{(c)}$, one can equivalently in this
situation consider $\xi$ as the limit of points in either one of the domains
$C_{\rightarrow}$, $ C_{\leftarrow}$ of $C^{(c)}$.
The advantage of the latter is the
derivation, as a by-product, of the analyticity with respect to $\xi$ in the domains
$C_{\leftrightarrows}$
of the transforms $\tilde f_{\leftrightarrows, \ell}(\xi)$ thus obtained.

It is convenient to use here the parametrization $z= z[\theta,\Psi]$ (see Eq. (\ref{param})) of $X^{(c)}$
which,  for $\xi \in \gamma_0^{(c)}$,
yields the following expression of $[z\cdot \xi]$:
\begin{equation}
[z[\theta,\Psi]\cdot \xi(\phi + i \eta)] = \sinh \Psi - \cosh \Psi \cos (\theta - \phi - i \eta).
\label{chiralzxi}
\end{equation}
As shown by this expression, for any fixed complex values of $\xi$ ($\xi \in
C_{\leftrightarrows}$),
$\Psi$ and ${\rm Im}\theta$, the complex point
$[z[\theta,\Psi]\cdot \xi(\phi + i \eta)]$ varies on an ellipse parametrized by ${\rm Re}\theta$.
For $z=x\in X$ (i.e. $\Psi$ and $\theta$ real), this ellipse encloses the origin;
this shows the necessity of defining the FH-kernel with $s$ integer.
These considerations lead one to the following

\vskip 0.5cm
\noindent
{\bf Proposition-Definition 3}
\ {\sl Given a function $f$ in ${\mathcal H}_{(reg)}(X)$,
we define its \ \ {\rm chiral Fourier-Helgason transforms}  as
the following two sequences of functions:
$\{\tilde f_{\leftrightarrows, \ell}(\xi), $ $ \ell\  {\rm integer}\  \geq 0 \}$ :
\begin{equation}
\tilde{f}_{\leftrightarrows, \ell}(\xi ) =
\int_X [x\cdot \xi]^{-\ell -1}
f(x)\ {\rm d}\sigma(x) ;
\label{Fourierchi}
\end{equation}
for each $\ell$, the two functions
$\tilde{f}_{\rightarrow, \ell}(\xi ) $ and
$\tilde{f}_{\leftarrow, \ell}(\xi )$ are defined by the integral at the r.h.s. as holomorphic functions in the respective
domains
$C_{\rightarrow}$
and $C_{\leftarrow}$ of $C^{(c)}$.

Moreover,
for every function $ f \in H^2_{(reg)}({\mathcal T}_{\rightarrow})$
(resp. $H^2_{(reg)}({\mathcal T}_{\leftarrow}) $), there is a {\rm unique} (non-vanishing) chiral FH-transform which is
$\{\tilde{f}_{\rightarrow, \ell}(\xi )\} $
(resp. $\{\tilde{f}_{\leftarrow, \ell}(\xi )\}$). }

\vskip 0.4cm

The proof of the latter statement, namely the fact that the integral (\ref{Fourierchi})
vanishes e.g. for $\xi \in
C_{\rightarrow}$ if $f $ belongs to
$H^2_{(reg)}({\mathcal T}_{\leftarrow}) $,
is proved by contour-distortion in the complex $\theta-$plane (making use of the parametrization
(\ref{param}) of $X^{(c)}$ and of
(\ref{chiralzxi})): this is an exact counterpart of the result of Proposition 8 for the Lorentzian case.
Similarly, one would also prove that the chiral FH-transforms both vanish for all the functions $f$
which belong to
$H^2_{(reg)}({\mathcal T}^{\pm}) $ (i.e. the analog of Proposition 16).

\subsection{Inversion of the transformation}

In order to prove the analog of Theorem 2 for the case of the chiral FH-transformation,
we need to define for each point $z$ in
${\mathcal T}_{\rightarrow}$
(resp. ${\mathcal T}_{\leftarrow} $)
an appropriate class of relative cycles
$\gamma(z)$ in $H^1(C^{(c)}, \{\xi;\ [z\cdot \xi] =0 \})$ with support contained in
$C_{\rightarrow}$ (resp. $ C_{\leftarrow}$): the end-points of this support will respectively belong
to the two linear generatrices of the cone $C^{(c)}$ defined by the equation $[z\cdot \xi] =0$.
In view of Proposition 5, it is sufficient to define $\gamma(z)$ for $z \in h_{\rightarrow}$
(resp. $h_{\leftarrow}$), i.e. of the form $z= z_v = (0,i \sinh v, \cosh v),\  v>0$ (resp.
$v<0$).
In that case, we specify the cycle $\gamma(z_v)$ in the manifold
$\gamma_0^{(c)}= \{\xi =\xi(\Phi) =  (1, \sin \Phi, \cos \Phi);\  \Phi =\phi + i \eta \}$
as follows:  since $[z\cdot \xi] = - \cos(\Phi -iv)$ vanishes at $ \Phi = \pm \frac{\pi}{2} + iv$, we
choose $\gamma(z_v)$ as the path $\phi  \to \Phi= \phi +iv$,  with $\phi$  increasing from
$-\frac{\pi}{2}$
to $\frac{\pi}{2}$: according to the sign of $v$,
the support of $\gamma(z_v) $ belongs either to $C_{\rightarrow}$
or to $C_{\leftarrow}$.
For an arbitrary point $z$ in
${\mathcal T}_{\rightarrow}$
(resp. ${\mathcal T}_{\leftarrow} $), which is
obtained (in view of proposition 5) by the action of a certain transformation $g$ of
$G$ on an appropriate point $z_v$, the corresponding cycle $\gamma(z)$ is defined by the action of $g$
on $\gamma(z_v)$, its support being
contained in the corresponding one-dimensional complex manifold
$\gamma^{(c)} = g \gamma_0^{(c)}$.
This is satisfactory since the domains
$C_{\rightarrow}$
and $C_{\leftarrow}$ as well as the set with equation
$[z\cdot \xi] =0$ are invariant under the action of $G$.
Moreover, by considering any path in the group $G$ as a union of arbitrarily small
successive paths, one easily sees that in the above construction the cycle $\gamma(z)$ remains
for all $z$ in a continuously varying relative homology class
in $H^1(C^{(c)}, \{\xi;\ [z\cdot \xi] =0 \})$
\cite{[Le]} (since the support of $\gamma(z)$
can be distorted by a succession of small homotopies keeping its end-points respectively in
the generatrices of $C^{(c)}$ with equation $[z\cdot \xi]=0$).

The following geometrical property will play an important role:
\begin{proposition}
Let $z\in
{\mathcal T}_{\rightarrow}$ and $z'$ in the closure of
${\mathcal T}_{\leftarrow} $.
Then for all $\xi \in \ {\rm supp}\ \gamma(z)$ (with the above definition
of $\gamma(z)$), one has $[z'\cdot \xi] \neq 0 $.
\end{proposition}
\begin{proof}\ \ Using again the $G-$invariance properties of
${\mathcal T}_{\leftrightarrows}$, we can take $z=z_v$ and $\xi \in \gamma(z_v)$, while
$z'$ is represented as follows (in view
of (\ref{param}) and (\ref{Tleft})):
$$ z'=(\sinh(\psi +i\varphi),\
\cosh(\psi +i\varphi) \sin(u+iv'),\
\cosh(\psi +i\varphi) \cos(u+iv')) $$
with
$\tanh  v' \leq -\frac{|  \sin \varphi|}{\cosh\psi}.$

We then have (since $v>0$):\ \
$\tanh (v'- v) < -\frac{|  \sin \varphi|}{\cosh\psi},$\
\ \ which implies:
\begin{equation}
\cosh (v-v') > \frac{\cosh \psi}{(\cosh^2 \psi- \sin^2 \varphi)^{1 \over 2}}
\ \ \ {\rm or} \ \ \ \  e^{\psi} + e^{-\psi} < 2 \cosh (v-v') |\cosh(\psi +i\varphi)|.\
\label{ellipse}
\end{equation}
On the other hand, one has:
$$ [z'\cdot \xi] = \sinh(\psi +i \varphi) -
\cosh(\psi +i \varphi) \cos (u- \phi +i(v'-v)), \ \ \ \  - {\pi \over 2} \leq \phi \leq {\pi
\over2},$$
which shows that the complex point $[z'\cdot \xi]$ varies on an ellipse
whose major axis is equal to
$ 2a = 2 \cosh (v-v') |\cosh(\psi +i\varphi)| $
and whose foci are the points $F= e^{\psi +i \varphi}$
and $F'= e^{-(\psi +i \varphi)}$.
Then the inequality (\ref{ellipse}) means that $OF + OF' <  2a$, namely that
the origin is always strictly inside the ellipse described by the point $[z'\cdot \xi]$
and therefore the statement
$[z'\cdot \xi] \neq 0 $ is proved .
\end{proof}

\noindent
Using the relative cycles $\gamma(z)$ previously defined,
we can now prove the following analog of Theorem 2:

\vskip 0.4cm
\noindent
{\bf Theorem 4}
{\sl Let $f(x)$ belong to
$H^2_{(reg)}\left({\mathcal T}_{\leftrightarrows}\right)$
and let
$\tilde{f}_{\leftrightarrows, \ell}(\xi ) $
be its chiral FH-transform.
Then the holomorphic function $F(z)$ in
${\mathcal T}_{\leftrightarrows}$
whose boundary value is $f$ is given in terms of
$\tilde{f}_{\leftrightarrows, \ell} $
by the following formula:
\begin{equation}
F(z) =  \frac{1}{4\pi^2}
\sum_{\ell = 0}^{\infty}
(2 \ell +1)
\int_ {\gamma(z)}
[z\cdot \xi]^{\ell} \
\tilde{f}_{\leftrightarrows, \ell}(\xi ) \
{\rm d}\mu_{ \gamma^{(c)}} (\xi)
\label{invfourierchi}
\end{equation}
In the latter,
${\rm d}\mu_{\gamma^{(c)}}$ is the restriction to $\gamma^{(c)}$ of the holomorphic extension
of the form $[i_{\Xi} \omega](\xi)$ (see 4.2) to $C^{(c)}$.}

\vskip 0.4cm
One notices that the integral over $\xi$ at the r.h.s. of Eq. (\ref{invfourierchi}) is well-defined
and holomorphic in
${\mathcal T}_{\leftrightarrows}$
since the function
$\tilde{f}_{\leftrightarrows, \ell}(\xi) $ is holomorphic in the corresponding domain
$C_{\leftrightarrows}$ of $C^{(c)}$ and therefore integrable on $\gamma(z)$,
by construction of the latter.
As a matter of fact, in view of this analyticity property,
the corresponding integral on the real cycle $\gamma$ (used for the Lorentzian inversion formula
(\ref{invfourier})) would give a vanishing integral in the present case.

In order to prove Theorem 4,
one  makes use of the
Cauchy-type representation (\ref{Cauchy}) for the case of the tubes ${\mathcal T}_{\leftrightarrows}$
together with the following expression of the Cauchy kernel on $X^{(c)}$
whose proof is given below:
\begin{proposition}
The Cauchy kernel on $X^{(c)}$ is given by the following double sum-integral
\begin{equation}
\frac{1}{(z'-z)^2} =   \frac{1}{4}
\sum_{\ell = 0}^{\infty}
(2 \ell +1)
\int_ {\gamma(z)}
[z\cdot \xi]^{\ell} \
[\xi\cdot z']^{-\ell -1}
{\rm d}\mu_{ \gamma^{(c)}} (\xi)
\label{expcauchychi}
\end{equation}
which is absolutely convergent for $(z,z')$ in
${\mathcal T}_{\rightarrow}
\times {\mathcal T}_{\leftarrow} $
and in ${\mathcal T}_{\leftarrow}
\times {\mathcal T}_{\rightarrow}. $
This formula remains meaningful when one of the points
is taken real, e.g. $z'=x$, provided the limit is taken from the appropriate tuboid.
\end{proposition}

If we plug the expression (\ref{expcauchychi}) of $\left[(x-z)^2\right]^{-1}$ into
the r.h.s. of (\ref{Cauchy}), invert the integrals over $x$ and over $(\ell,\xi)$
and take into account the defining formula
(\ref{Fourierchi}) of
$\tilde f_{\leftrightarrows, \ell}(\xi) $,
we readily obtain formula (\ref{invfourierchi})
for both types of configurations .
So, proving Theorem 4  amounts to proving Proposition 19.

\subsection{Chiral invariant perikernels: a new representation for the second-kind Legendre functions and
the Cauchy kernel on $X^{(c)}$}

We now introduce a class of $G^{(c)}-$ invariant kernels on $X^{(c)}$ which play the same role as
the perikernels of subsection 4.3 with respect to the chiral tuboids.
\footnote
{In dimension $d$, $(d \geq 2),$ such kernels arise in the context of quantum field theory on
a $d-$dimensional quadric of signature $(+,+,-,\cdots,-)$,
interpreted as a $d-$dimensional anti-de Sitter spacetime manifold.}

\vskip5pt Such a ``chiral  invariant perikernel'' is a holomorphic
function ${\rm W}(z,z')$ holomorphic in the ``cut-domain''
$\Delta_{\chi} = \{(z,z')\in X^{(c)}\times X^{(c)}; \ [z\cdot
z']\in \Theta_{\chi} = {\mathbb C}\setminus [-1,+1]\}$ and
satisfying (\ref{covcom}). It then follows from Proposition 6 that
${\rm W}$ is holomorphic in particular in the two tuboids
${\mathcal T}_{\rightarrow} \times {\mathcal T}_{\leftarrow}$ and
${\mathcal T}_{\leftarrow} \times {\mathcal T}_{\rightarrow}$ of
$X^{(c)}\times X^{(c)}$. In view of its $ G-$invariance property
(\ref{covcom}), ${\rm W}(z,z')$ can also be identified with a
function (i.e. its reduced form) ${\rm w}([z\cdot z'])$
holomorphic in the cut-plane $ \Theta_{\chi}$.

This provides a second class of Lorentz invariant kernels ${\mathcal W} (x, x')$ on $X$,
obtained as the boundary value of a holomorphic function ${\rm W}\left(
z,z'\right) $ from the analyticity domain
${\mathcal T}_{\leftarrow }\times {\mathcal T}_{\rightarrow }$ (or its opposite).
\vskip 5pt

Basic examples of such chiral invariant perikernels
\footnote{In the application to anti-de Sitter quantum field theory, these examples are interpreted
as the two-point functions (or ``propagators'') of massive Klein-Gordon-like fields \cite{[BBMS]}
up to an appropriate normalization.}
are provided by the second-kind Legendre functions,
as shown by the following statement

\begin{proposition}
The following integral representation holds for $(z,z')$ varying
in the pair of domains
${\mathcal T}^{r,l} = {\mathcal T}_{\rightarrow }\times {\mathcal T}_{\leftarrow }$ and
${\mathcal T}^{l,r} = {\mathcal T}_{\leftarrow }\times {\mathcal T}_{\rightarrow }$:
\begin{equation}
Q_{\ell}([z\cdot z']) = (-1)^{\ell+1} \frac{1}{2}
\int_ {\gamma(z)}
[z\cdot \xi]^{\ell} \
[\xi\cdot z']^{-\ell -1}
{\rm d}\mu_{ \gamma^{(c)}} (\xi)
\label{legendre2}
\end{equation}
The latter integral is still convergent when one of the points, e.g. $z'$, is taken
as a limiting real point from either tuboid
$ {\mathcal T}_{\rightarrow }$ or ${\mathcal T}_{\leftarrow }$,
$z$ remaining inside the opposite tuboid.
It moreover defines the second-kind Legendre functions $Q_{\ell}$  as a
family of chiral invariant
perikernels
depending on the integer $\ell$.
\end{proposition}
\begin{proof}\ \
The independence of the integral (\ref{legendre2}) with respect to the choice of
the cycle $\gamma(z)$ inside its homology class is a consequence of proposition 10 and of Stokes theorem,
since $\gamma(z)$ is a relative cycle  whose boundary belongs to an analytic set \cite{[Le]},
namely the set with
equation $[z\cdot \xi]=0$.
Now the fact that this integral defines a holomorphic function of $(z,z')$ in the product
tuboid
${\mathcal T}^{r,l}$ (and in its opposite)
including also the limiting configurations with $z'$ real (as described in the statement)
is a direct consequence of Proposition 18, since
in that domain the
factor $[\xi\cdot z']^{-\ell -1}$ never becomes singular for $\xi$ varying in $\gamma(z)$.
Moreover, since the chosen representatives $\gamma(z)$ are precisely distorted
by the action of the group $G$ on $\gamma(z_v)$ when $z$ varies in its tuboid, this implies
the invariance property of the integral
(\ref{legendre2})
under all the transformations $(z,z') \to (gz,gz');\ g\in G$
(accompanied by the allowed change $\gamma(z) \to g \gamma(z)$).
The function defined by this integral is therefore a function of the invariant $[z\cdot z']$,
holomorphic in the image $\Theta_{\chi}$ of
${\mathcal T}^{r,l}$

Let $ {\rm W}^{r,l}_{(\ell)}(z,z')$ be the  holomorphic function defined by
the integral (\ref{legendre2})
in the tuboid
${\mathcal T}^{r,l}$.
In order to compute
$ {\rm W}^{r,l}_{(\ell)}(z,z') =
{\rm w}^{r,l}_{(\ell)} ([z\cdot z'])$
explicitly,
we choose
a simple configuration in the border of
${\mathcal T}^{r,l}$,
namely the pair $(z,z') = (z_v, b)$,
which is such that $[z\cdot z'] = -\cosh v$.
The corresponding cycle $\gamma (z_v)$ in the manifold
$\gamma_0^{(c)}$ has been defined at the beginning of subsection 6.2.
We then obtain (by going successively from the integration variable $\phi$
which parametrizes $\gamma (z_v)$ to
$t= \tan \phi$):
$${\rm w}^{r,l}_{(\ell)} ([z\cdot z']) =
(-1)^{\ell+1} \frac{1}{2 }
\int_{-\frac{\pi}{2} }^{\frac{\pi}{2}}
\cos^{-(\ell +1)} (\phi + i v) \ \cos^{\ell} \phi\  {\rm d}\phi $$
\begin{equation}
= (-1)^{\ell+1} \frac{1}{2 }
\int_{-\infty}^{\infty} (\cosh v -i t \sinh v)^{-(\ell +1)} \frac {{\rm d}t}{(1+t^2)^{\frac{1}{2}}}
\label{leg2}
\end{equation}
By contour distortion to $t= i\tau;\  \tau \geq 1$ and putting $\tau = \cosh u$, one gets
\begin{equation}
{\rm w}^{r,l}_{(\ell)} ([z\cdot z']) =
(-1)^{\ell+1}
\int_0^{\infty} (\cosh v +\cosh u \sinh v)^{-(\ell +1)}\  {\rm d}u
=  Q_{\ell}([z\cdot z'])
\label{Leg2}
\end{equation}
A similar computation would yield the same result by taking configurations $(z,z')$ in the opposite
tuboid
${\mathcal T}^{l,r}$.
\end{proof}

\begin{remark}
The function ${\rm W}_{(\ell)}(z,z')$ is a solution in both
variables of the Laplace-Beltrami equation $[\Delta_X + \ell (\ell
+1)] {\rm W}_{(\ell)}(z,z') =0$ on the manifold $X^{(c)}$ (or in
other words, the Klein-Gordon equation on complexified anti-de
Sitter spacetime). For such a $G-$invariant solution, this
equation reduces to a Legendre equation in the variable $[z\cdot
z'] $ and the required analyticity domain $\Theta_{\chi}= {\mathbb
C} \setminus [-1,+1]$ selects the second-kind Legendre function
$Q_{\ell}([z\cdot z'])$ as the unique relevant solution (up to a
constant factor).
\end{remark}

We now come back to the Cauchy kernel $[(z-z')^2]^{-1} = \{-2([z\cdot z']+1)\}^{ -1} $,
whose analyticity in $\Theta_{\chi}$ and
$G-$invariance show that it also defines a {\sl chiral} invariant perikernel on $X^{(c)}$:
we prove that the r.h.s. of Eq. (\ref{expcauchychi}) is precisely a representation of this
perikernel in the pair of tuboids
${\mathcal T}^{r,l}$ and ${\mathcal T}^{l,r}$.

\vskip 0.2cm
\noindent
{\bf Proof of Proposition 19}
\vskip 0.4cm
\noindent
As in the case of Proposition 11, one could give a complete
computation exhibiting a Cauchy-Fantappi\'e form of the kernel (proof B)), but
for brevity we shall only give the fast proof, based
on the previous integral representation (\ref{legendre2}) of the second-kind
Legendre functions
and
on the following well-known formula \cite{[ER]}, valid for all $Z \in \Theta_{\chi}$:
\begin{equation}
\frac{1}{Z-1} = \sum_{\ell=0}^{\infty} (2 \ell +1) Q_{\ell}(Z)
\label{cauchydiscret}
\end{equation}
By plugging the expression (\ref{legendre2}) of
$Q_{\ell}(-[z\cdot z'] )
= (-1)^{\ell +1}
Q_{\ell}([z\cdot z'])$
into the r.h.s. of (\ref{cauchydiscret}) and taking into account the
integrability conditions specified in Proposition 20,
one readily obtains
(\ref{expcauchychi}) with the specifications of Proposition 19.

\section{Conclusion:
Fourier-Helgason representation of the holomorphic decomposition in the tuboids }

Being given any function $f(x)$ in ${\mathcal H}_X$, it admits a
decomposition of the form $f = f^+  + f^-  + f_\rightarrow  +
f_\leftarrow$, where each function is the boundary value on $X$ of
a holomorphic function in the corresponding tuboid ${\mathcal
T}^+, {\mathcal T}^-, {\mathcal T}_\rightarrow, {\mathcal
T}_\leftarrow $ of $X^{(c)}$. This decomposition of $f$ has been
obtained in section 3 by applying to $f$ a Cauchy kernel acting as
a projection operator onto the Hardy space of each of the four
tuboids. The results of section 4, in particular Proposition 8 and
Theorem 2, completed by Proposition 16, show that in this
decomposition the first two components $f^{+}$ and $f^-$ are
completely characterized respectively by the Lorentzian
Fourier-Helgason transforms $\tilde f_{-,\nu}(\xi)$ and  $\tilde
f_{+,\nu}(\xi)$ of $f$. Each of these transforms lives on
$C^+_{(\xi)} \times {{\mathbb R}_+}_{(\nu)}$, where $\nu$ labels
the principal series of irreducible representations of
$SO_0(1,2)$. Formulae (\ref{Fourier}) and (\ref{invfourier}) then
imply the following {\sl Plancherel formulae} which introduce
$L^2-$isomorphisms between the Hardy spaces $H^2({\mathcal
T}^{\pm})$ and the appropriate spaces of FH-transforms on
$C^+_{(\xi)} \times {{\mathbb R}_+}_{(\nu)}$ (note that equivalent
formulae making use of the range $\nu \in ]-\infty,0]$ instead of
$[0,+\infty[$ for the FH-transforms are also valid) .

For any pair of functions $(f,g)$
in ${\mathcal H}(X)$ and
their corresponding decompositions, one has:
\begin{equation}
\int_X \overline {f^-(x)} \  g^-(x)\  {\rm d}\sigma(x) =
\frac{1}{2\pi^2} \int_0^{\infty} \frac{\nu \tanh \pi\nu}{e^{\pi \nu} \cosh \pi \nu} {\rm d}\nu
\int_{\gamma} \overline {\tilde f_{+,\nu} (\xi)}\  \tilde g_{+,\nu}(\xi)\ {\rm d}\mu_{\gamma}(\xi)
\label{Plancherel-}
\end{equation}
and
\begin{equation}
\int_X \overline {f^+(x)} \  g^+(x)\  {\rm d}\sigma(x) =
\frac{1}{2\pi^2} \int_0^{\infty} \frac{\nu \tanh \pi\nu}{e^{-\pi \nu} \cosh \pi \nu} {\rm d}\nu
\int_{\gamma} \overline {\tilde f_{-,\nu} (\xi)}\  \tilde g_{-,\nu}(\xi)\ {\rm d}\mu_{\gamma}(\xi)
\label{Plancherel+}
\end{equation}
Similarly, the results of section 6 (Proposition-Definition 3 and
Theorem 4) show that the other two components $f_{\rightarrow}$
and $f_{\leftarrow}$ of the decomposition of $f$ are completely
characterized respectively by the chiral Fourier-Helgason
transforms $\tilde{f}_{\rightarrow, \ell}(\xi ) $ and
$\tilde{f}_{\leftarrow, \ell}(\xi )$ of $f$.  These transforms
live respectively on $C_{\rightarrow} \times {\mathbb N}$ and
$C_{\leftarrow} \times {\mathbb N}$ and are associated with the
discrete series of irreducible representations of $SO_0(1,2)$.
Corresponding Plancherel formulae are expected to follow from
formulae (\ref{Fourierchi}) and (\ref{invfourierchi}).

\vskip 0.2cm In conclusion, we have given for the functions on the
one-sheeted hyperboloid an explicit treatment of the
Gelfand-Gindikin program, exhibiting in that case the construction
of a holomorphic decomposition into four tuboids and its identity
with an appropriately defined  Fourier-Helgason decomposition in
terms of the range of the corresponding irreducible
representations of $G$. Although more sophisticated, the result
remains very close to the holomorphic decomposition into four
tubes and the corresponding support decomposition in the Fourier
variables for the functions on ${\mathbb R}^2$.

\vskip10pt {\bf Acknowledgements.} Part of this work was carried
out at the Erwin Schr\"odinger Institute in Vienna, which the
authors would like to thank for hospitality and financial support.

\end{document}